\documentclass[apj]{emulateapj}

\newcommand\up{\it u{'}}
\newcommand\gp{\it g{'}}
\newcommand\rp{\it r{'}}
\newcommand\ip{\it i{'}}
\newcommand\zp{\it z{'}}
\received{November 4, 2017}
\revised{December 23, 2017}
\accepted{January 2, 2018 }

\shorttitle{NGC\,796: A deep visible AO study}
\shortauthors{Kalari et al.}

\begin{document}

\title{The Magellanic Bridge cluster NGC\,796: \\Deep optical AO imaging reveals the stellar content and initial mass function of a massive open cluster}

\email{venukalari@gmail.com}

\author{Venu M. Kalari}
\affil{Departamento de Astronomia, Universidad de Chile, Casilla 36-D, Santiago, Chile}
\author{Giovanni Carraro}
\affil{Dipartimento di Fisica e Astronomia Galileo Galilei, Vicolo Osservatorio 3, I-35122, Padova, Italy}
\author{Christopher J. Evans}
\affil{UK Astronomy Technology Centre, Royal Observatory Edinburgh, Blackford Hill, Edinburgh EH9 3HJ, UK}
\author{Monica Rubio}
\affil{Departamento de Astronomia, Universidad de Chile, Casilla 36-D, Santiago, Chile}



\begin{abstract}

NGC\,796 is a massive young cluster located 59\,kpc from us in the diffuse intergalactic medium of the 1/5--1/10\,$Z_{\sun}$ Magellanic Bridge, allowing to probe variations in star formation and stellar evolution processes as a function of metallicity in a resolved fashion, providing a link between resolved studies of nearby solar-metallicity and unresolved distant metal-poor clusters located in high-redshift galaxies. In this paper, we present adaptive optics {\it{gri}}H$\alpha$ imaging of NGC\,796 (at 0.5$\arcsec$, which is $\sim$\,0.14\,pc at the cluster distance) along with optical spectroscopy of two bright members to quantify the cluster properties. Our aim is to explore if star formation and stellar evolution varies as a function of metallicity by comparing the properties of NGC\,796 to higher metallicity clusters. We find from isochronal fitting of the cluster main sequence in the colour-magnitude diagram an age of 20$^{+12}_{-5}$\,Myr. Based on the cluster luminosity function, we derive a top-heavy stellar initial mass function (IMF) with a slope $\alpha$\,=\,1.99$\pm$0.2, hinting at an metallicity and/or environmental dependence of the IMF which may lead to a top-heavy IMF in the early Universe. Study of the H$\alpha$ emission line stars reveals that Classical Be stars constitute a higher fraction of the total B-type stars when compared with similar clusters at greater metallicity, providing some support to the chemically homogeneous theory of stellar evolution. Overall, NGC\,796 has a total estimated mass of 990$\pm200$\,$M_{\odot}$, and a core radius of 1.4$\pm$0.3\,pc which classifies it as a massive young open cluster, unique in the diffuse interstellar medium of the Magellanic Bridge.
\end{abstract}




\section{Introduction}

\begin{figure*}
\includegraphics[width=2.2\columnwidth]{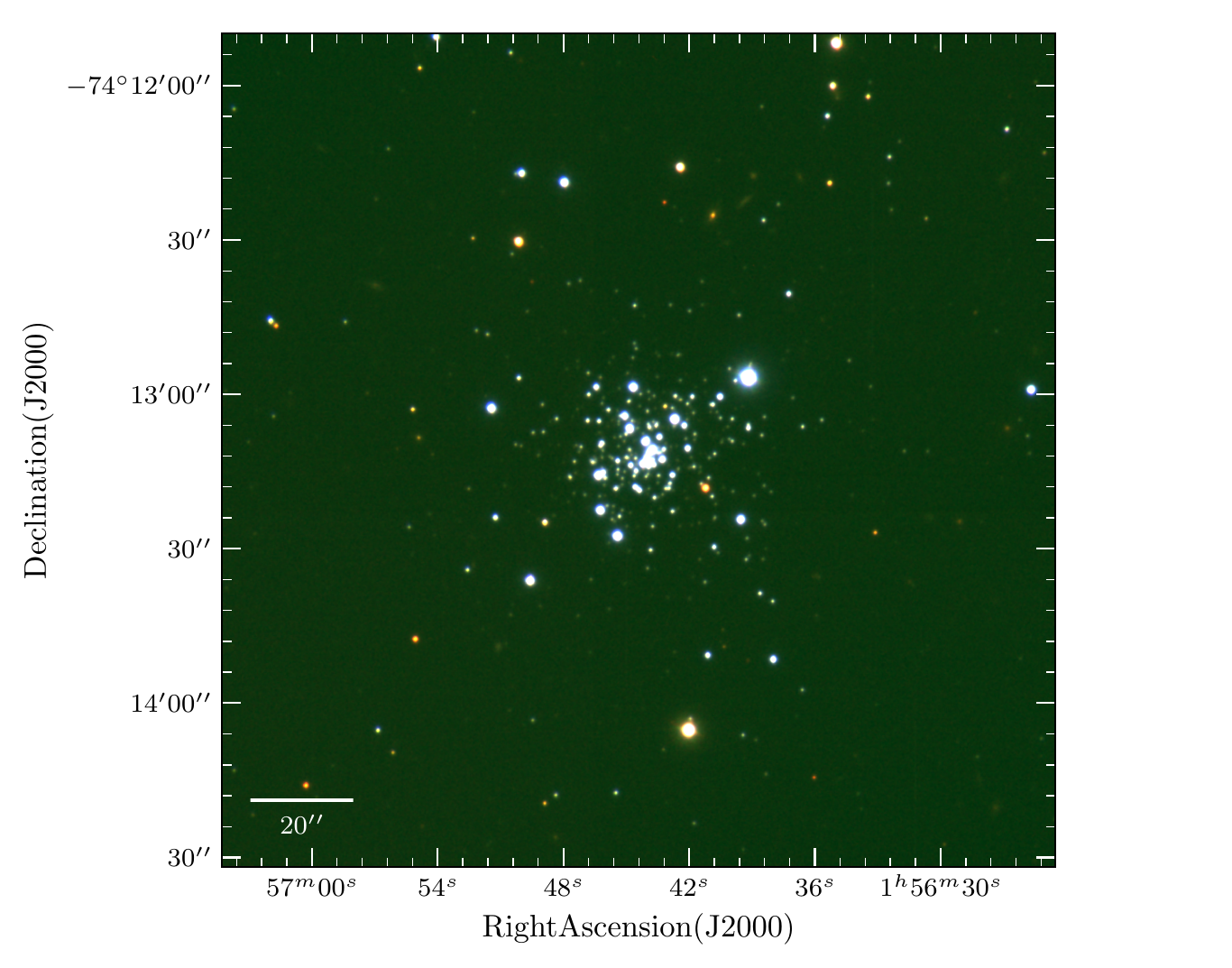}
\caption{NGC\,796 as viewed in the SOAR-AO 3$\arcmin \times 3\arcmin$ three-colour igmage, where {\it r/g/b} is from the $\ip/\rp/\gp$ 50s images, respectively. North is up and east is to the left. The lower-left scalebar gives 20$\arcsec$, which at the distance to NGC\,796 subtends a distance of $\sim$6\,pc.\label{fig:f1}}
\end{figure*}

Young open clusters constitute vital laboratories for confronting theories on star formation and stellar evolution processes, as they contain both massive stars as well as those arriving on the main sequence. Over the past century owing to technological advancements, observational studies of open clusters in the Milky Way have permitted astronomers to test theories of star formation and stellar evolution enabling us to form a picture of how stars form and evolve (Iben 2013). Extending that picture to metal-poor young open clusters is essential, as metallicity ($Z$) is one of the two primary parameters (the other being stellar mass) governing star formation and stellar evolution (Bromm \& Larson 2004). Deep and high angular resolution observations facilitated by the modern generation of instruments have over the past two decades allowed to explore in detail open clusters in the Magellanic Clouds, which are the nearest metal-poor galaxies (comprised of the Large and Small Magellanic Cloud at 1/2 and 1/5\,$Z_{\odot}$ at 50 and 61\,kpc respectively; see D'Onghia \& Fox 2016) where individual stars can be resolved. These studies have instigated debates on key topics; e.g., whether the stellar initial mass function (IMF) varies as a function of $Z$ (see Schneider et al. 2018.; Andersen et al. 2009; Kroupa 2002; Sirianni et al. 2000); does stellar evolution proceed differently at much lower $Z$ (e.g. Hunter et al. 2007; Mokiem et al. 2007; Maeder et al. 1999); or is pre-main sequence evolution different at lower $Z$ (see Kalari \& Vink 2015; De Marchi et al. 2011). Studies of metal-poor young open clusters (and particularly those at 1/5\,$Z_{\odot}$, or lower) will provide empirical evidence to test current theories of star formation and stellar evolution as a function of $Z$, allowing us to place observational constraints on the same in the high-redshift early Universe. {\footnote {Based on observations of a large sample of star-forming disc galaxies, Kewley \& Kobulnicky (2005, 2007) conclude that on average a 0.15\,dex decrease in $Z$ corresponds to a unit redshift, so open clusters at 1/5\,$Z_{\odot}$ or lower correspond to a redshift of 4 or lower, before the peak epoch of star formation, and the galaxy assembly era. } }

With this aim, we have undertaken a program to observe and characterise stars clusters in the Magellanic Bridge (Irwin et al. 1990) using deep and high angular resolution photometry, and spectroscopy. The Magellanic Bridge is a stream of H\,{\scriptsize I} that connects the 1/2\,$Z_{\odot}$ Large Magellanic Cloud (LMC) and the 1/5\,$Z_{\odot}$ Small Magellanic Cloud (SMC). In our initial study, we focus on the most interesting target, NGC\,796. NGC\,796 (alias L\,115, ESO\,30-6 or WG71\,Cl* 9) is a dense cluster (Nishiyama et al. 2007; Piatti et al. 2007; Ahumada et al. 2009; Bica et al. 2015) that lies $\sim$4000\,pc east of the SMC wing. NGC\,796 lies in a region of higher column density within the inter-cloud region. A molecular gas cloud of mass $\sim$ 10$^3$\,$M_{\odot}$ has been detected $\sim$60\,pc to the south of the cluster (Mizuno et al. 2006). The metallicity estimated from stellar abundances of B-type stars in the Wing region near the cluster is around 1/5\,$Z_{\odot}$ (Lee et al. 2005), while both gas-phase and stellar abundances in the enveloping inter-cloud region suggest a lower metallicity of 1/10\,$Z_{\odot}$ (Rolleston et al. 1999; Dufton et al. 2008; Lehner et al. 2008). As the metallicity of stars within the cluster has hitherto been unmeasured, for the remainder of this work we assume NGC\,796 $Z$\,=\,1/5\,$Z_{\odot}$, although we caution that the metallicity may be even lower. In either scenario, the low $Z$, distance (in between that of the SMC and LMC, at $\sim$55-60\,kpc; see D'Onghia \& Fox 2016), and near isolation makes NGC\,796 remarkable in many aspects. Additionally, of all the known star clusters in the Magellanic Bridge (Bica \& Schmitt 1995), NGC\,796 is the most compact and dense as yet identified and is hence an ideal target to study numerous facets of resolved star formation and stellar evolution at low $Z$.

There have been three previous optical studies of NGC\,796. Ahumada et al. (2009, 2002) report from integrated stellar spectra properties of clusters in the Magellanic Bridge. Their results are based mainly on the Balmer line strengths and the continuum of the integrated spectra compared to models. For NGC\,796, they report an age of 20\,Myr, with a mean reddening $E$($B-V$) of 0.03 and a distance comparable to the SMC. Results from Washington photometry of individual stars in the cluster by Piatti et al. (2007) obtain a similar reddening (assuming a comparable distance to our derived value), but an age of 110\,Myr. In contrast, results from $BV$ photometry of individual stars by Bica et al. (2015) find a reddening between  0.03--0.04, but a distance of only $\sim$41\,kpc (implying the cluster is closer to us than the LMC) and an age of $\sim$40\,Myr.

In this study, we present new high angular resolution ($\sim$0.5$\arcsec$, which translates to 0.14\,pc at the adopted cluster distance) adaptive optics (AO) deep optical broadband and H$\alpha$ imaging of NGC\,796, along with optical spectroscopy of two bright members. Our data is an improvement over the literature in terms of angular resolution (as we can resolve the cluster centre), and because our photometry is deeper than previous studies, allowing for the first time to study the high and intermediate mass stellar content, the cluster properties, star-formation history, and the IMF at such low-$Z$ in the Magellanic Bridge. This paper is organized as follows. In Section 2, the data used in this study are described. The main results emerging from the study are presented in Section 3. A discussion on the stellar IMF, fraction of classical Be stars (cBe), and the overall cluster properties is made in Section 4. Finally, in Section 5 a summary of the work is presented.

\section{Data}
\subsection{Optical adaptive optics imaging}
NGC\,796 was observed using the SAM (SOAR Adaptive Optics module) imager mounted on the 4.1-m SOAR (Southern Astrophyiscal Research Observatory) telescope located at Cerro Pach{\'o}n, Chile. Adaptive optics is necessary to achieve ground-based angular resolutions around 0.5$\arcsec$ or better, essential to resolve centrally concentrated clusters such as NGC\,796 at the distance to the Magellanic Clouds. The SAM imager covers a 3 arcmin sq. field on the sky with a single 4096$\times$4112 e2v CCD at a pixel scale of 45.4\,{\it mas}. The CCD has a noise of 3.8\,{$e^-$}, and a gain of 2.1 $e^-$/ADU. The detector becomes saturated around 65000 counts. Ultraviolet (UV) light from a pulsed laser at 355\,nm is used to create an artificial laser guide star to measure wavefront distortion caused by low-altitude turbulence (up to a distance of few km in the lower atmosphere) with a Shack-Hartmann sensor. A deformable mirror compensates for the distortion at optical wavelengths and projects this corrected image onto the camera resulting in an AO corrected image. The angular resolution at optical wavelengths is improved, typically by a few tenths of an arcsec from the natural seeing depending on the wavelength. The science observations are separated from the laser beam with a dichroic at 370\,nm, and observations at shorter wavelengths are not possible. Stars outside the main field are used to provide tip-tilt guiding using a fast tertiary mirror. 

We utilized the $\gp\rp\ip$ filters along with a H$\alpha$ filter for our observations. These filters were preferred over the standard {\it BVRI} set as there are significant leaks in the $B$ bandpass. These filters resemble closely the United States Naval Observatory (USNO) $\up\gp\rp\ip\zp$ filter set (Smith et al. 2002), rather than the Sloan Digital Sky Survey (SDSS) {\it ugriz} filters as seen from the filter response curves{\footnote {http://www.ctio.noao.edu/soar/content/filters-available-soar}}. The H$\alpha$ filter was used primarily to identify any H$\alpha$ emission line stars (for e.g. cBe stars, or Herbig AeBe stars). The H$\alpha$ filter has a Full Width Half Maximum (FWHM) of 65\AA, and is centred on 6561\AA.

\begin{table}
	\centering
	\caption{SOAR AO imaging observing log}
	\label{tab:example_table}
\begin{tabular}{cccc}        
\hline\hline               
Filter & Exp. Time & Delivered FWHM \\
& (s) & (arcsec) \\
\hline
$\gp$ & 1 & 0.69$\arcsec$\\
$\gp$ & 10 & 0.74$\arcsec$\\
$\gp$ & 50 & 0.72$\arcsec$\\
$\gp$ & 100 & 0.75$\arcsec$\\
$\gp$ & 200 & 0.72$\arcsec$\\
$\rp$ & 1 & 0.45$\arcsec$\\
$\rp$ & 10 & 0.45$\arcsec$\\
$\rp$ & 50 & 0.47$\arcsec$\\
$\rp$ & 100 & 0.46$\arcsec$\\
$\rp$ & 200 & 0.5$\arcsec$\\
$\ip$ & 1 & 0.43$\arcsec$\\
$\ip$ & 10 & 0.46$\arcsec$\\
$\ip$ & 50 & 0.47$\arcsec$\\
$\ip$ & 100 & 0.47$\arcsec$\\
$\ip$ & 200 & 0.5$\arcsec$\\
H$\alpha$ & 60 & 0.41$\arcsec$\\
H$\alpha$ & 120 & 0.44$\arcsec$\\
H$\alpha$ & 200 & 0.58$\arcsec$\\
H$\alpha$ & 240 & 0.57$\arcsec$\\

\hline                        
      
\end{tabular}\\
\end{table}

Observations were conducted in closed loop during the night of 10 October 2016, centred on $\alpha\,=\,01^{h}56^{m}44^{s}$, $\delta\,=\,-74\degr13\arcmin12\arcsec$. Atmospheric conditions were stable with a clear, cloudless night. The median uncorrected seeing was $\sim$0.9--1$\arcsec$. Images in the $\gp\rp\ip$ filters were taken with exposure times of 1, 10, 50, 100, 200s to provide a large dynamic magnitude range without saturating the bright stars. In H$\alpha$, we took exposures of 60, 120, 200 and 240s. The airmass throughout our observations was between 1.39--1.42.  The final image quality (angular resolution) varied for each filter, and is tabulated in Table 1. The image quality is given by the final FWHM as measured from bright isolated stars on the images using the {\textsc{iraf}}{\footnote{{\textsc{iraf}} is distributed by the National Optical Astronomy Observatory, which is operated by the Association of Universities for Research in Astronomy (AURA) under a cooperative agreement with the National Science Foundation.}} task {\it imexam}. This accounts for the compensation provided by the AO. For $\rp\ip$H$\alpha$, the image FWHM is between 0.4$\arcsec$-0.5$\arcsec$. For $\gp$, the FWHM is slightly worse at $\sim$0.7$\arcsec$. Assuming a distance of 59\,kpc (see Section\,3.4), our images are able to resolve stars further than 0.14\,pc apart. 

Flats and bias images were obtained before the observing run. Bias subtraction, flat fielding, and bad pixel masks were applied to the corrected raw images using the {\textsc{iraf}} task {\it ccdproc}. Accurate astrometric solutions to our data proved challenging as no resolved stars were found towards the centre of the cluster in conventional astrometric catalogues, instead we utilized the USNO v2 catalogue of Monet et al. (2002). An accurate solution was obtained correcting for distortion and isoplanatic effects using the {\textsc{iraf}} task {\it ccmap} in the tangent-plane projection with a third order polynomial fit. Astrometric residuals are around 0.2$\arcsec$ or better, as measured from stars outside the cluster centre. This value gives the lower limit to our astrometric accuracy. Given that the distortion effects are minimal towards the centre of the Field of View (FoV), we expect the astrometric accuracy to be similar or better towards the cluster centre, although we are unable to quantify it given the lack of astrometric standard stars in that location. The final reduced images (of the 50s exposures) are shown as a three-colour image in Fig. 1. The image quality and depth of our photometry may be judged by the resolved stars in the cluster centre, and the detection of faint cluster galaxies. 

 
We obtained $\gp\rp\ip$ observations of 21 standard stars from Smith et al. (2002) during twilight, mid-night and in the morning. The standard stars span an airmass range of 0.9 allowing for an accurate calibration of the magnitude zero point and airmass effects. We converted the USNO $\gp\rp\ip$ standard star system to the oft-used SDSS standard {\it gri} system using the equations detailed by the SDSS survey{\footnote {http://classic.sdss.org/dr7/algorithms/sdssphot.ps}}. Our choice was driven primarily by ease of comparison with standard stellar model sets. In addition, we included standard star fields at low and high airmasses from VPHAS+ (VLT Photometric H$\alpha$ survey of the Southern Galactic Plane and Bulge; Drew et al. 2014), to calibrate the variation of airmass and colour using the SDSS {\it gri}H$\alpha$ magnitudes of the survey. We performed aperture photometry on the standard stars using the {\textsc{iraf}} task {\it daophot} to obtain precise instrumental magnitudes. The aperture size chosen was equal to the FWHM of each image. 

To transform the instrumental magnitudes to the standard SDSS system, we fit functions to derive the magnitude zeropoints ($\zeta$), extinction terms ($k_{\chi}$), and colour terms ($k$) for each broadband filter using an iterative orthogonal fitting function. A weighted fit (with the weight being the inverse of the magnitude errors in both x and y) was applied. We fit equations of the form
\begin{equation}
g_{\rm STD} = g_{\rm inst} + \zeta_{g} + \chi \times k_{\chi, g} +(g-r) \times k_{g}
\end{equation}
\begin{equation}
r_{\rm STD} =  r_{\rm inst} + \zeta_{r} + \chi \times k_{\chi, r} +(g-r) \times k_{r}
\end{equation}
\begin{equation}
i_{\rm STD} = i_{\rm inst} + \zeta_{i} + \chi \times k_{\chi, i} +(g-i) \times k_{i}
\end{equation}
where {\it gri} standard and instrumental magnitudes are identified by the STD and inst subscript respectively. $\chi$ is the airmass in the respective observations. The rms deviations of the fits with the standards were 0.03, 0.04, 0.07 mag in {\it gri} filters, respectively, which are propagated into the final magnitude errors of the observations. The derived values for {\it gri} filters are, respectively, $\zeta$\,=\,26.096, 26.208, 25.8175, $k_{\chi}$\,=\,$-$0.195, $-$0.115, $-$0.05, and $k$\,=\,0.115, $-$0.0169, 0.05. We calibrated our H$\alpha$ magnitudes after calculation of the $r$-band values, employing standards from the VPHAS$+$ survey.
 
\begin{figure}
\centering
\includegraphics[width=\columnwidth]{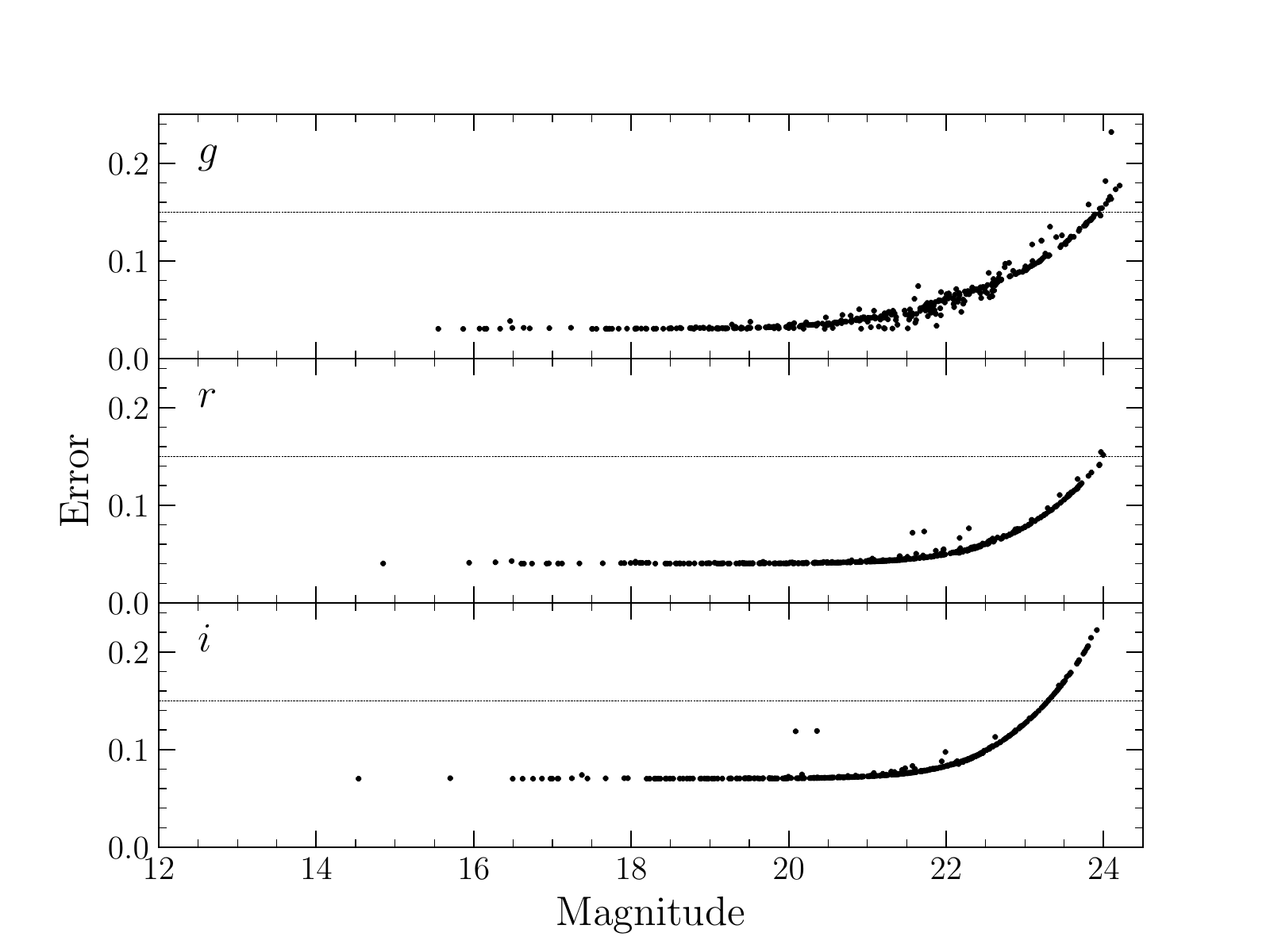}
\caption{Random uncertainties of calibrated magnitudes as a function of magnitude, where the top, middle and bottom panels give the values for the $g$, $r$ and $i$-band {\it psf} photometry,  respectively. The solid line in each panel marks the cut-off for the error criterion, where $\sigma$\,=\,0.15, above which the photometry was not incorporated into the analysis.\label{fig:error}}
\end{figure}

For analysis of the cluster images we employed the {\textsc {starfinder}} Interactive Data Language ({\textsc {IDL}}) package of Diolaiti et al. (2000) to detect sources and compute their photometry. Compared with standard point spread function ({\it psf}) fitting suites, {\textsc {starfinder}} provides comparable or better results in the isoplanatic case (e.g. Diolaiti et al. 2000) from a locally determined {\it psf}. We used {\textsc {starfinder}} to determine the empirical local {\it psf} from isolated bright sources throughout the FoV, while propagating the Gaussian photon noise, sky variance and fitting residuals to estimate the photometric errors.

The empirical {\it psf} was constructed using (at least three) bright and isolated stars throughout the FoV. These were found in each image individually, after which any nearby fainter sources were removed by hand. Sources were then detected in each image, after which the {\it psf} was re-estimated accounting for any new faint detections. The source detection was then run again, from which we compiled a list of sources and their instrumental magnitudes for each filter and exposure time. Sources above the saturation limit of 65000 counts, below the 3$\sigma$ detection threshold, or having a {\it psf} correlation coefficient less than 75\% were discarded. Note that the correlation coefficient is a measure of the similarity to the empirical {\it psf}. Finally, we concatenated the resulting source lists per filter (577 sources were detected in $r$), removing any multiple detections within 0.1$\arcsec$ and cross-matching multiple filters based on the $r$-band astrometry. The resulting band-merged catalogue is given in Table. 2. In total, we detect 429 stars in the {\it gri} filters.

\begin{figure}
\centering
\includegraphics[width=0.9\columnwidth]{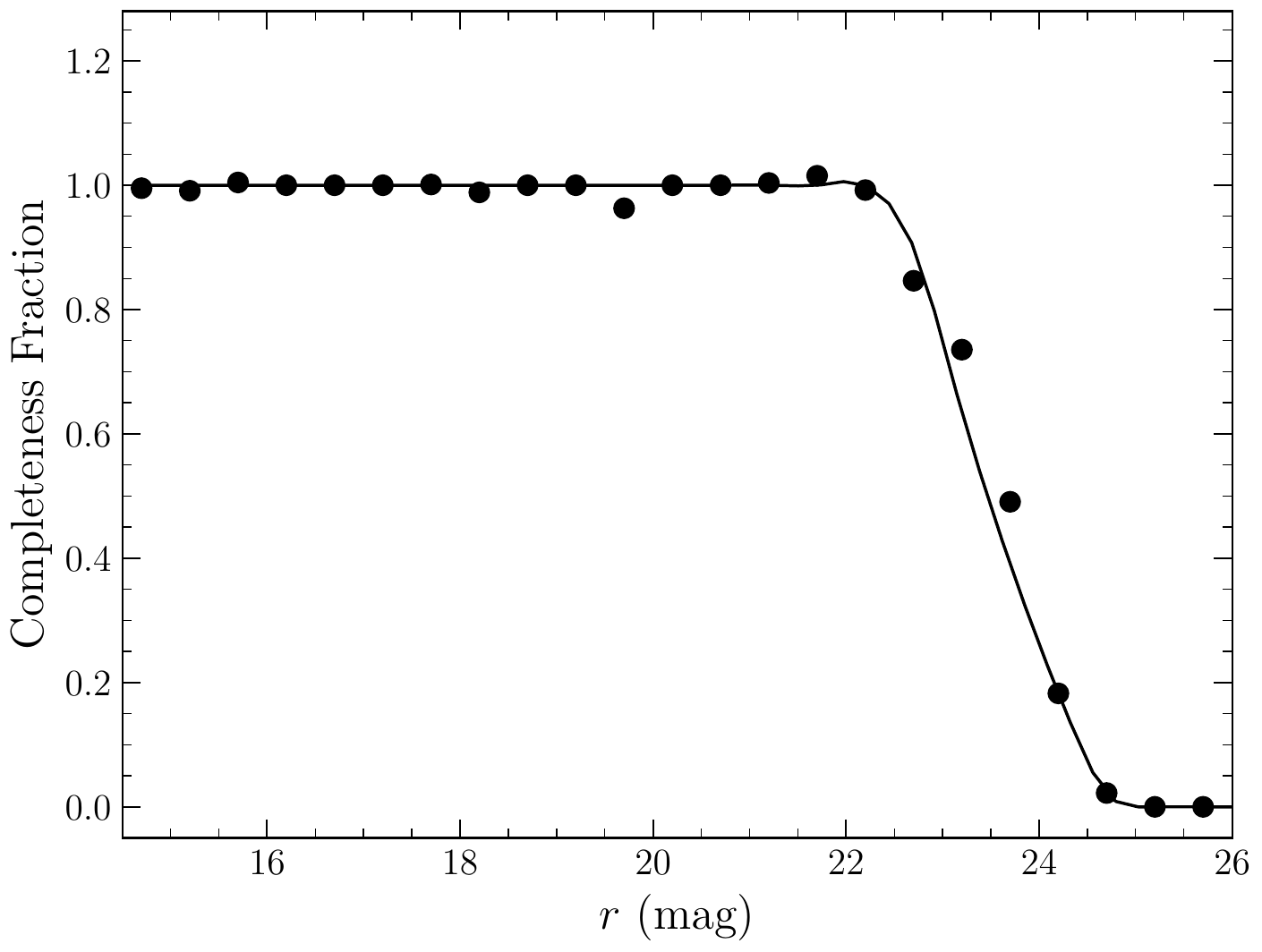}
\caption{Completeness limit of our $r$-band photometry as a function of magnitude within the central 1.5$\arcmin$ region of the FoV. \label{complete}}
\end{figure}

In Fig. 2, we compare the resulting photometric errors as a function of measured magnitude. The formal uncertainties on our photometry are small and within 0.1\,mag even for most of our faintest stars. We impose a cut-off in magnitude uncertainty of 0.15\,mag in each of our filters for the data used in further analysis. To study the completeness of our photometry, we conducted artificial star experiments using the {\textsc{iraf}} task {\it addstar}. Around 5\% (of the total detected sources in each image at each magnitude bin of 1\,mag) were randomly added within the central 1.5$\arcmin$ (which completely includes the central cluster region visible in Fig. 1). We then ran {\textsc {starfinder}} again and computed the fraction of recovered stars (i.e. the number of newly added stars recovered with {\textsc {starfinder}} to the total number of new stars added). This experiment was repeated 100 times and used to estimate the average recovery fraction. The recovered fraction is nearly 100\% until 21\,mag in each filter, after which it drops sharply. The 50\% completeness level is reached for $g$ at 22.2\,mag, $r$ at 23.5\,mag and in $i$ at 23.4\,mag. The results of the completeness for the $r$ filter are shown in Fig.\,\ref{complete}

\begin{table*}
	\centering
	\caption{$gri$ photometry of all stars. Only ten lines are shown here to display the form and content, but the full table is available in the online version of this journal, or electronically via remote ftp. }
	\label{tab:example_table}
\begin{tabular}{lcccccl}        
\hline\hline                
 ID & Right Ascension ($\alpha$) & Declination ($\delta$) & $g$ & $r$ & $i$ & Remarks  \\
& J2000 & J2000 & (mag) & (mag) & (mag)  & \\
\hline
  1 & $01^{h}56^{m}24^{s}53$  & $-74\degr14\arcmin20\arcsec5$  & 21.86$\pm$0.04 & 21.71$\pm$0.04 & 21.72$\pm$0.07  & \\
  167 & $01^{h}56^{m}41^{s}71$ & $-74\degr13\arcmin10\arcsec3$  & 24.53$\pm$0.03 & 23.97$\pm$0.13 & 23.84$\pm$0.16  & \\
  209 & $01^{h}56^{m}42^{s}83$ & $-74\degr12\arcmin14\arcsec9$  & 17.69$\pm$0.03 & 17.07$\pm$0.04 & 16.75$\pm$0.07  & Non-member based on CMD \\
   &  &   &  &  &   & $J$=15.37,\,$H$=14.84,\,$K$s=14.66 $^{a}$ \\
  212 & $01^{h}56^{m}42^{s}94$ & $-74\degr13\arcmin03\arcsec9$  & 15.93$\pm$0.03 & 16.24$\pm$0.04 & 16.5$\pm$0.07  & H$\alpha$ emission line and HBe$^{b}$ candidate \\
   &  &   &  &  &   & $J$=16.13,\,$H$=16.12,\,$K$s=15.89 $^{a}$ \\
  237 & $01^{h}56^{m}43^{s}51$ & $-74\degr13\arcmin11\arcsec6$  & 17.19$\pm$0.03 & 17.52$\pm$0.04 & 17.78$\pm$0.07  & H$\alpha$ emission line candidate \\
  262 & $01^{h}56^{m}43^{s}97$ & $-74\degr13\arcmin09\arcsec9$  & 16.07$\pm$0.03 & 16.48$\pm$0.04 & 16.86$\pm$0.07  & B1-B3\,V \\
  283 & $01^{h}56^{m}44^{s}34$ & $-74\degr13\arcmin08\arcsec3$  & 16.81$\pm$0.04 & 17.26$\pm$0.04 & 17.55$\pm$0.07  & B2\,V \\
  376 & $01^{h}56^{m}46^{s}30$ & $-74\degr13\arcmin14\arcsec3$  & 18.05$\pm$0.03 & 18.40$\pm$0.04 & 18.70$\pm$0.07  & H$\alpha$ emission line candidate \\
  404 & $01^{h}56^{m}47^{s}06$ & $-74\degr12\arcmin55\arcsec1$  & 20.46$\pm$0.03 & 20.62$\pm$0.04 & 20.76$\pm$0.07  & \\
  440 & $01^{h}56^{m}49^{s}06$ & $-74\degr13\arcmin24\arcsec2$  & 18.79$\pm$0.03 & 18.49$\pm$0.04 & 18.29$\pm$0.07  &  Non-member based on CMD \\
  \hline                        
\end{tabular}\\
\text\raggedleft{(a) Magnitudes from Kato et al. (2008); (b) According to Nishiyama et al. (2008) }
\end{table*}


\subsection{Optical low-resolution spectroscopy}

Optical spectroscopy of two cluster members was obtained during an observing run with the Bollers \& Chivens Spectrograph mounted on the 2.5-m du Pont telescope atop the Las Campanas Observatory in Las Campanas, Chile. The two stars were selected based on their position within the cluster centre and brightness which indicate that they are most probably cluster members. The 300\,l/mm grating was used providing a spectral dispersion of 3\,\AA/pixel in a 1$\arcsec$ wide slit. The grating angle of the spectrograph was set to 10.8$^{\rm \circ}$, covering the $\lambda\lambda$\,3650-6670\AA~wavelength range. 

\begin{figure}
\centering
\includegraphics[width=0.9\columnwidth]{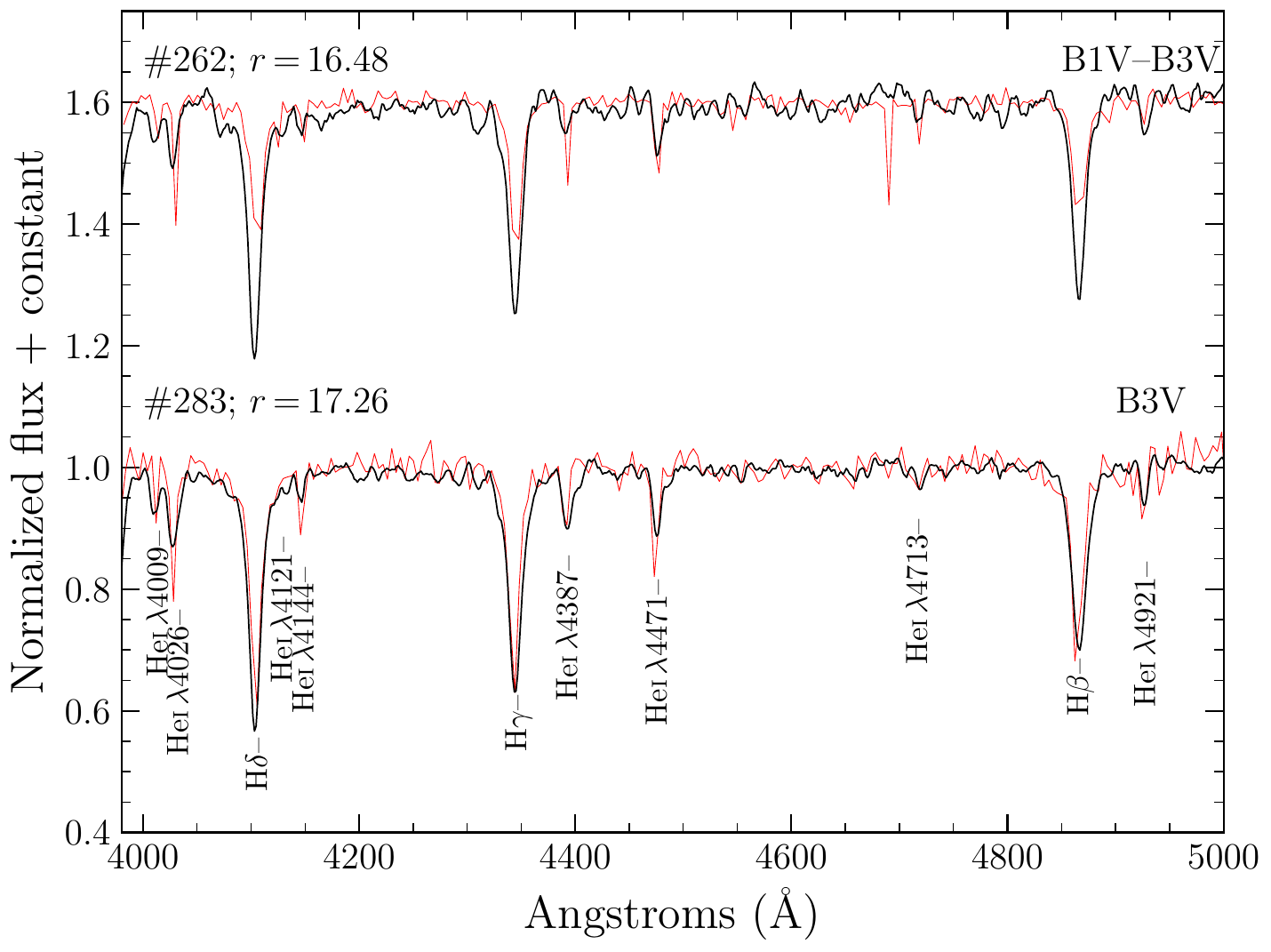}
\caption{Normalized spectra of \#262 (top) and \#283 (bottom) across the MK classification wavelength regime shown in solid black lines. Spectra are vertically offset by 0.6 continuum units for display, with the main spectral lines marked. Overplotted in solid red are smoothed high-resolution spectra of isolated stars of spectral type B2\,V and B3\,V in the SMC cluster NGC\,346. \label{spectra}}
\end{figure}

The spectrum was obtained on May 5$^{\rm th}$ 2017 towards the end of the night in photometric conditions with a seeing of 0.6$\arcsec$ but at an airmass of 1.8. The slit was rotated to a position angle of 173$^{\rm \circ}$, so as to place two stars (\#262 and 283 from Table 2) on the slit. The total on source exposure time was 1000s. Dome and twilight flats, and biases with the same setup were taken at the end of the night and used to reduce the spectrum using the {\textsc{iraf}} task {\it ccdproc}. NeHeAr arc lamp spectra were taken immediately before and after the science observation and used to wavelength calibrate the spectrum with the {\textsc{iraf}} task {\it identify}. The Bollers \& Chivens spectrograph does not have an atmospheric dispersion corrector so, given the non-parallactic slit angle used to observe the two stars, we did not attempt to flux calibrate our data. 1D spectra of the two stars were extracted from the final wavelength-corrected reduced 2D spectrum, while correcting for sky lines using a window of 20 pixels on either side of the peak flux of each spectrum. The final reduced and wavelength-calibrated spectra, corrected to the barycentric rest frame, are shown in Fig.\,\ref{spectra}.

\section{Results}

\subsection{Cluster Morphology}

To provide quantitative statistics on the cluster morphology we first defined the cluster centre. Density maps of the region were created based on the coordinates from the $r$-band photometry. To avoid detection of non-physical structures due to binning, we employed a non-parametric kernel estimate approach to derive density maps. The adopted Kernel method does not make a priori assumptions on the data distribution, but smooths the contribution of each point (which is equally weighted over a local neighbourhood described by the Silverman Kernel estimator). The density maps created were binned into three magnitude ranges spanning 13 $<r<$ 24, 13 $<r<$ 18, and 18 $<r<$ 24 corresponding to all our sources, the brightest, and faintest, respectively. The densest position of the map is $\alpha$\,=\,$01^{h}56^{m}44^{s}$, $\delta$\,=\,$-74\degr13\arcmin05\arcsec$, and is taken as the nominal cluster centre. This point does not differ by more than 2$\arcsec$ between the three maps.

\begin{figure}
\centering
\includegraphics[width=0.9\columnwidth]{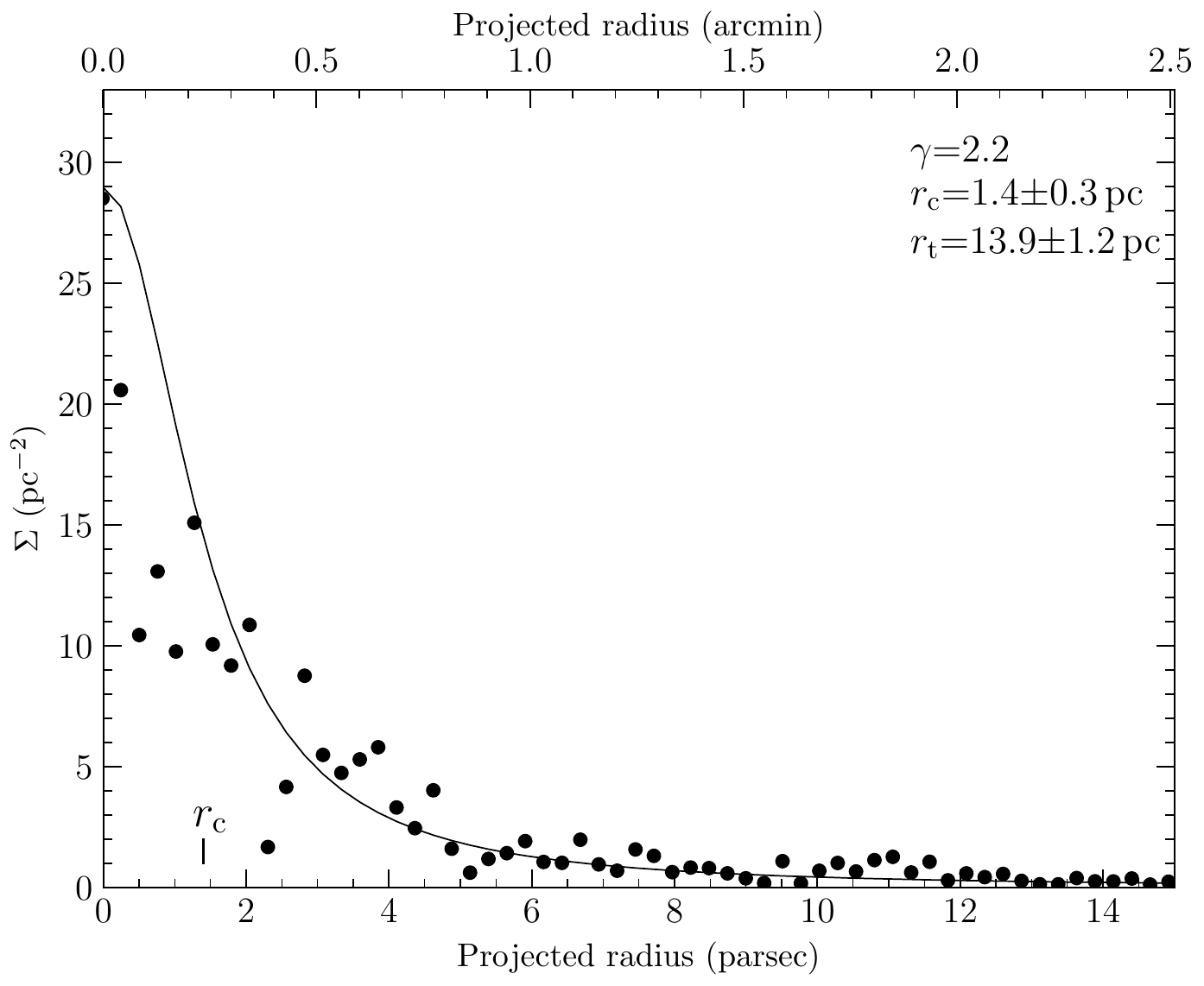}
\includegraphics[width=0.7\columnwidth]{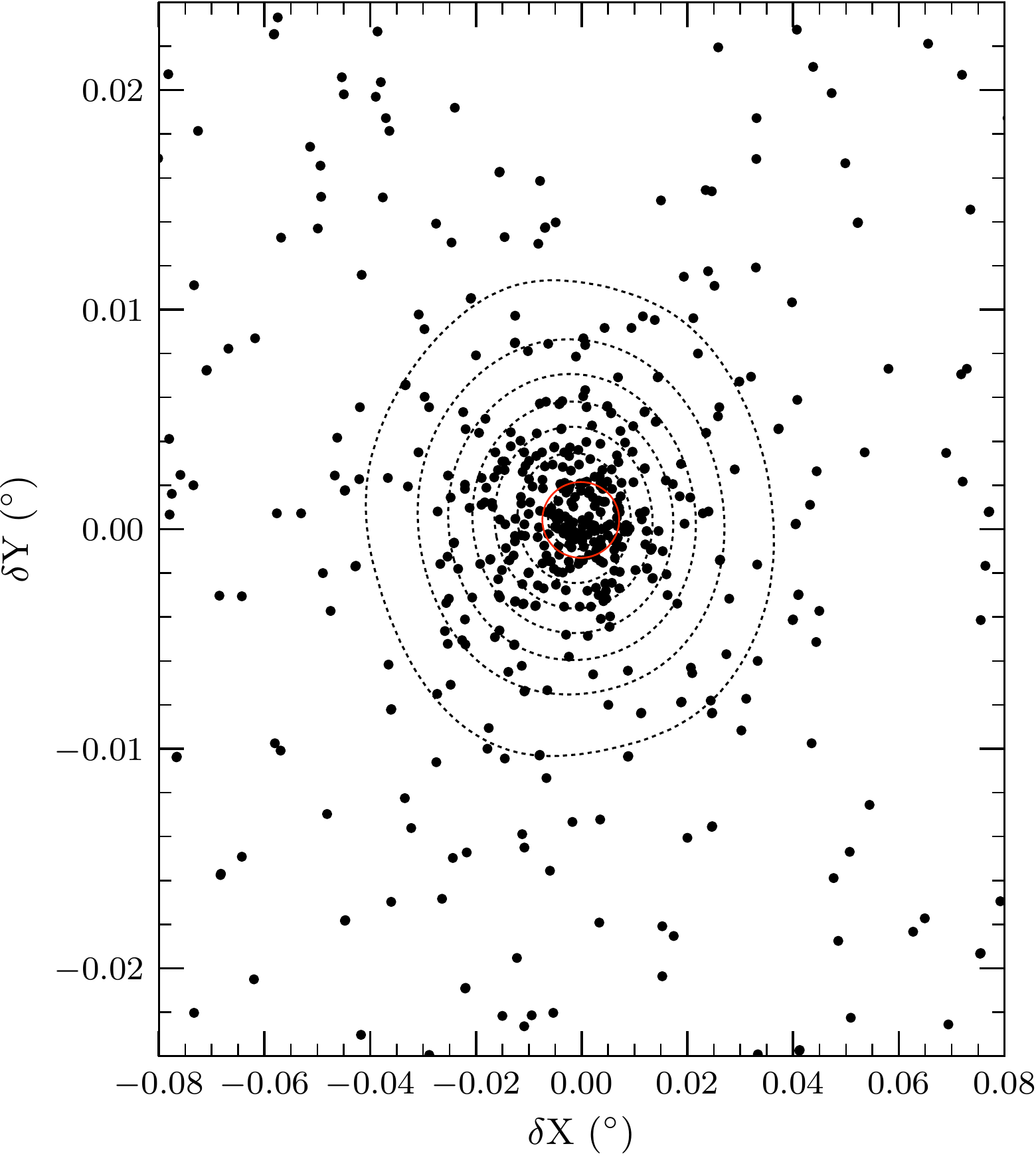}
\caption{Top: Surface density distribution of all sources with $r$-band photometry as a function of radius. The best-fit cluster profile is given by the solid line, and the resulting $r_{\rm c}$ marked. The best-fit parameters are given in the upper right corner. Bottom: Cluster finding chart of the inner region relative to the centre, with the density contours overplotted. Also shown is the core radius by the red circle. \label{surface}}
\end{figure}

\begin{figure*}
\plottwo{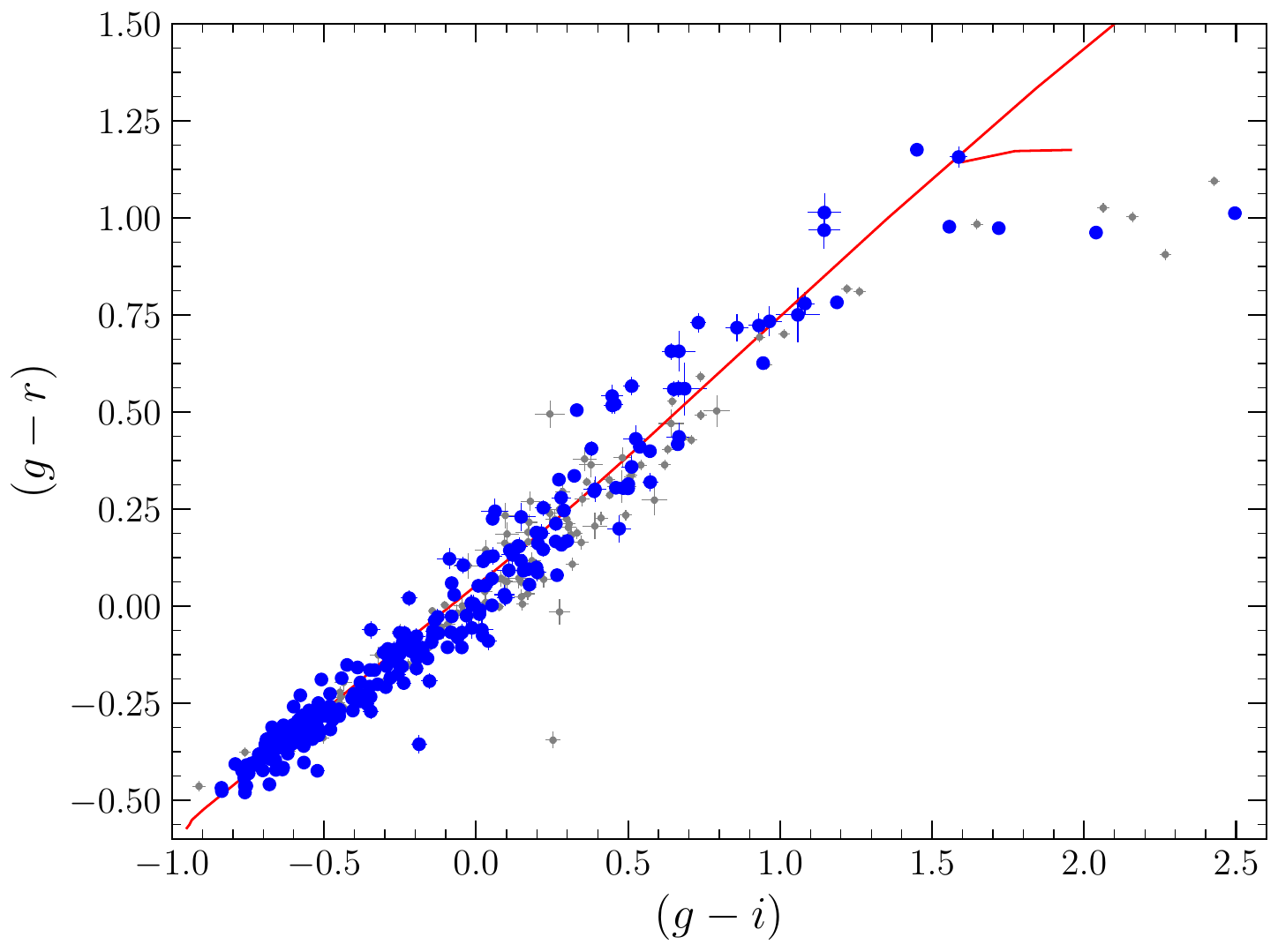}{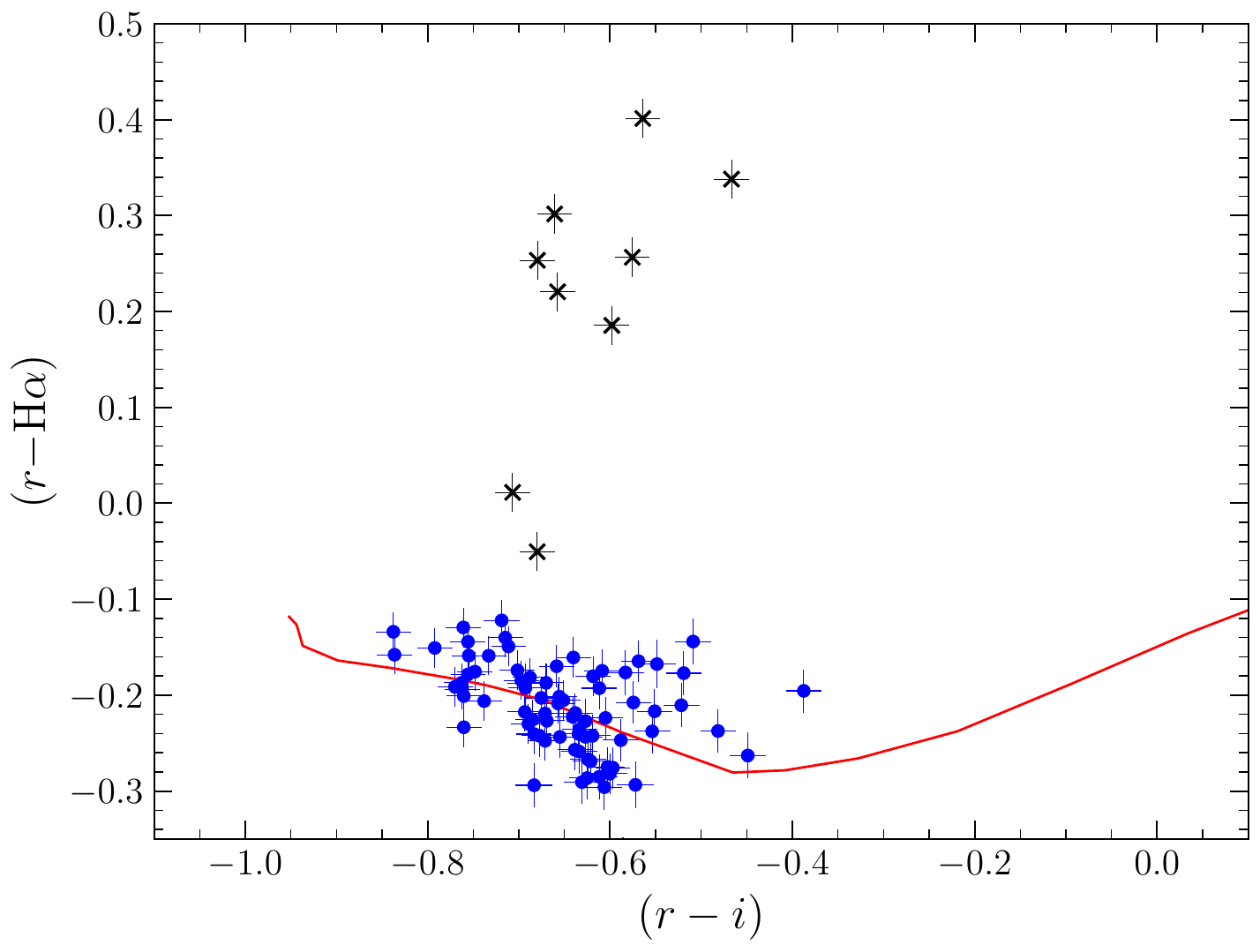}
\caption{Left: ($g-i$) vs. ($g-r$) two colour diagram of all stars falling with the cluster tidal radius (blue dots) and those outside (grey dots). The locus of 1/5\,$Z_{\odot}$ dwarfs from Castelli \& Kurucz (2003) is shown as a solid red line, with divergence at ($g-i$)\,=\,1.6 for the giants. The dwarf locus is shifted by $A_V$\,=\,0.1\,mag assuming $R_V$\,=\,2.75. Right: Blue dots here account only for stars with ($r-$H$\alpha$) random photometric uncertainties $<$0.1\,mag. Stars with marked H$\alpha$ excesses are shown as black asterisks. }
\label{ccd}
\end{figure*}

Based on the derived cluster centre, we compute the radial profile of the stellar surface density. Fig.\,\ref{surface} shows the distribution of sources as a function of projected radii (in units of arcmin and parsecs, assuming the distance to the cluster is 59\,kpc). We fit the radial profile of stellar surface density with a normalized surface density distribution given by
\begin{equation}
\Sigma = \Sigma_{0}[1+(r/a)^2]^{-\gamma/2}+\phi,
\end{equation}
which according to Elson et al. (1987) is ideally suited to model the distribution of young open clusters. Here $\Sigma$ is the stellar surface density, $\Sigma_{0}$ is the stellar surface density at the cluster centre, $r$ the radius, $a$ is the ratio of the cluster core to tidal radius, $\gamma$ is the power-law slope at radius reaching infinity, and $\phi$ is the field density. We fit the profile to our data using a least-squares fitting algorithm, and found the best-fit profile to have a core radius $r_{\rm c}$, where the stellar surface density drops to half its peak value, of 1.4$\pm$0.3\,pc, with the cluster tidal radius ($r_{\rm t}$) to be 13.9$\pm$1.2\,pc for the exponent $\gamma$\,=\,2.2. We therefore, restrict our likely cluster candidates further by imposing on them a radius criterion of $r<$\,13.9\,pc. We find that the inner part (not the centre) does not fit the profile well, and displays some anomalies. Similar anomalies in the surface brightness profiles were found among SMC, LMC and M33 young clusters (e.g. see Mackey \& Gilmore 2003; Werchan \& Zaritsky 2011; San Roman et al. 2012). In the cluster finding chart (Fig.\,\ref{surface}) we find a slight under-density in the inner regions compared to a classical profile, and similarity to the ring clusters of Mackey \& Gilmore (2003). However it is unclear how much physical significance can be attached to these findings (e.g. San Roman et al. 2012), and whether these anomalies are the result of the physical conditions during sub-cluster merging, or due to the initial conditions at the time of formation (Elson 1991). Note that according to the Galactic stellar population models of Robin et al. (2003), the contamination of foreground Galactic stars within our magnitude detection limits towards NGC\,796 is extremely low at $\lesssim$5 stars/\,arcmin$^{-2}$. We therefore expect a significantly high fraction of stars to be members based on the location alone. The final cluster candidate members are shown in the two-colour diagrams (TCD) and colour-magnitude diagrams (CMD) shown in Figs.\,\ref{ccd} and \ref{cmdall} respectively.
 
\subsection{Extinction}
In Fig.\,\ref{ccd}, we present the ($g-i$) vs. ($g-r$) TCD of all stars. The ($g-i$) and ($g-r$) colours have the largest baseline afforded by our observations thus providing a more sensitive measure to small values of extinction than the conventional ($g-r$) vs. ($r-i$) diagram. Overplotted are the interpolated colours of main-sequence and giant stars at 1/5\,$Z_{\odot}$, calculated by folding the Castelli \& Kurucz (2003) models through the appropriate filter curves. We note two sequences of stars, the cluster locus dominated by stars falling within the clusters tidal radius, and a locus of a few highly reddened late type stars found near the giant locus, with roughly half being stars outside the tidal radius. 

To estimate the reddening $E$($B-V$) of the cluster, we fit the main-sequence locus to the locus of cluster stars (blue dots in Fig.\,\ref{ccd}). The absolute value of reddening determined is 0.035$\pm$0.01 mag. The quoted uncertainty is the formal fitting error to the main-sequence locus. This value is so low that a reliable determination of the reddening law ($R_{V}$), even by combining these colours with archival near-infrared photometry to afford a longer baseline, would likely be unreliable and is subject to individual peculiarities and small photometric errors, rather than being dominated by the global reddening law (Sung et al. 1997). We instead utilize the results of Gordon et al. (2003) for the Small Magellanic Cloud where $R_V$\,=\,2.75. We obtain then a value of of the absolute extinction in the visual $A_{V}$\,=\,0.1\,mag, corresponding to the colour excess $E(B-V)$ of 0.035. This value is similar to the values quoted by both Piatti et al. (2007) and Bica et al. (2015) who both estimated the cluster reddening by fitting main sequence isochrones to the cluster locus in the optical CMD. As we shall later see, this value also agrees well with spectroscopic measurements. For the remainder of the paper, we adopt $A_V$\,=\,0.1, and $R_V$\,=\,2.75 to correct for extinction.

\subsection{{\rm H}$\alpha$ excess stars}
The ($r-$H$\alpha$) colour is a measure of H$\alpha$ line strength relative to the $r$-band photospheric continuum. Most main-sequence stars do not have H$\alpha$ in emission. Modelling their colour at each spectral type provides a template against which any colour excess due to H$\alpha$ emission can be measured from the observed ($r-$H$\alpha$) colour (De Marchi et al. 2011; Barentsen et al. 2011; Kalari et al. 2015). 

The ($r-i$) colour is used as a proxy for the spectral type, so that the ($r-i$) vs. ($r-$H$\alpha$) TCD can be used to identify stars with H$\alpha$ excess (Fig.\,\ref{ccd}). We only select stars having random photometric uncertainties in their combined ($r-$H$\alpha$) colours smaller than 0.15\,mag, to discard any stars with poor quality H$\alpha$ photometry. The majority of the stars with high quality H$\alpha$ photometry have ($r-i$) colours $<-$0.1, that is they are A-type stars or earlier. Thus, the H$\alpha$ photometry is only able to identify early-type stars with H$\alpha$ emission and not any late-type stars. The early-type stars in our sample (early B-early A) span a small colour range of $\sim$0.25 in ($r-i$) colours, and the combined error due to the combination of random photometric noise and extinction uncertainty $\sim$\,0.1 mag, means that the uncertainty on the estimated spectral subtype is $\sim$5 subclasses. Only stars showing large excesses, comparable to H$\alpha$ line widths $<-$15\AA~(where the negative sign denotes emission) are selected as emission line stars. Ten stars are selected as high-confidence H$\alpha$ emission line candidate stars, and based on their ($r-i$) colour (all with ($r-i$)$<-$0.18) are likely cBe stars. We discuss the implication of this finding in Section\,4.2. Our selection criteria for H$\alpha$ emission-line objects should result in relatively secure candidates, regardless of their estimated spectral type. However, the H$\alpha$ emitters thus identified will represent only an upper limit, and the actual number of H$\alpha$ emitters with smaller equivalent widths, below our imposed cut-off may well be larger. To identify those stars confidently, at least low-resolution spectroscopy with adequate sky subtraction is essential.

\begin{figure*}
\includegraphics[width=2.1\columnwidth]{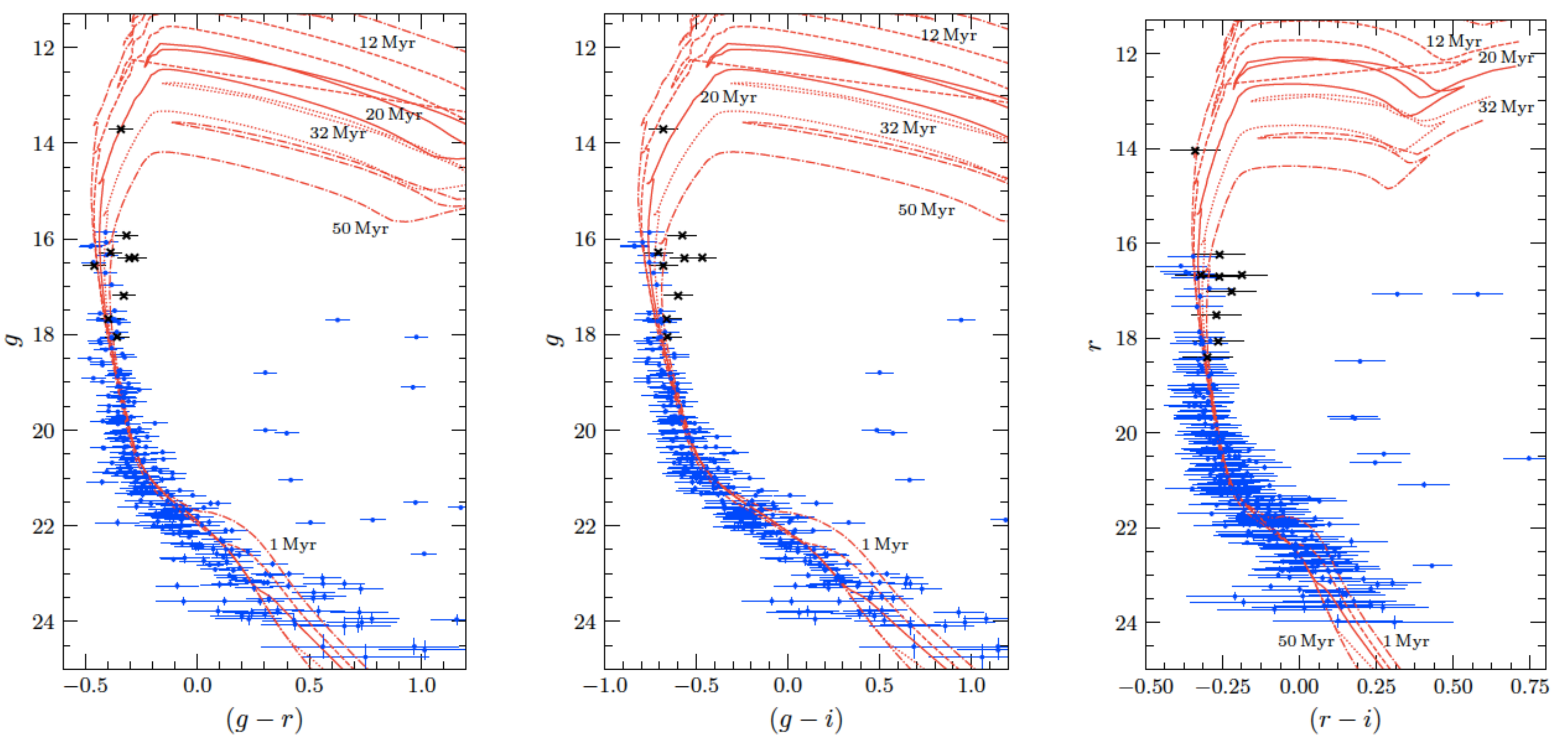}
\caption{Colour-magnitude diagrams for NGC\,796 in the ($g-r$) vs $g$, ($g-i$) vs. $g$ and ($r-i$) vs. $r$ planes are shown as blue dots in left, middle and right panels, respectively. Stars with H$\alpha$ excess are given by black asterisks. Overplotted are the 1/5\,$Z_{\odot}$ isochrones from Bressan et al. (2012) at ages of 1, 12, 20, 32 and 50\,Myr as dashed dotted, dashed, solid, dotted and dashed dotted red lines, respectively. The isochrones are shifted by an $A_V$ of 0.1\,mag assuming $R_V$\,=\,2.75, and by a distance modulus of 18.85\,mag.  }
\label{cmdall}
\end{figure*}

\subsection{Distance}

We plot different combinations of CMDs in Fig.\,\ref{cmdall}. In all CMDs a clear cluster locus on the zero age main sequence (ZAMS) is defined, with the stars lying between 14-24 magnitudes in $g$ (or in $r$). Overplotted are the 1/5\,$Z_{\odot}$ single star isochrones in the appropriate filters from Bressan et al. (2012). We show the isochrones for 1, 12, 20, 32 and 50\,Myr, corrected for extinction using the values determined in Section 3.2. The adopted distance was determined by fitting the main-sequence locus of each CMD to the zero age main-sequence locus defined by the isochrones. The fit was done between the ranges of $r$\,=\,16--21, and 18--21. We restricted our fit within this magnitude range, as the fainter cluster stars do not specify a well defined cluster sequence, but populate a comparatively broader space in the CMD and are not ideal to constrain the cluster distance (cf. Prisinzano et al. 2005). Hence we restrict our fit to the blue envelope of ZAMS stars lying within those magnitude ranges. The fit performed was a weighted $\chi^{2}$ minimization, where the weight applied was the inverse of the errors. The identified H$\alpha$ excess sources were excluded from the fit. Although the majority of red non-member candidates lying to the right of the locus formed by the points at $g$,($g-i$) of (16,0.2) to (24,1.0)  fall outside the tidal radius, we find a small fraction of stars in these positions falling inside the tidal radius, and outside the isochrones. They are most probably foreground giant stars. We also removed these stars from our fit because of their positions in the CMD.  

The resulting average of the distance modulus ($\mu$) from the main sequence fits to the three different CMDs is $\mu$\,=\,18.85$\pm$0.2\,mag, where the error is the difference between the fits to the different CMDs. The corresponding distance of $d$\,=\,59\,$\pm$0.8\,kpc places NGC\,796 between the determined distances to the LMC of 50\,kpc and to the SMC of 61\,kpc (D'Onghia \& Fox 2016), and closer to the SMC. This is expected given the proximity of the cluster to the wing of the SMC. If the Bridge feature is truly connected in space between the two Magellanic clouds then the distance to the Bridge from us may increase as we move closer towards the SMC. Note that our distance determination agrees within errors with the one utilized by Piatti et al. (2007) towards NGC\,796, but is much larger than the one determined by Bica et al. (2015) who find that NGC 796 lies much closer than the LMC to us at 40\,kpc. However, from spectroscopic parallax it is highly unlikely that NGC 796 lies much closer than 54\,kpc (see Section 3.6).

\subsection{Age}

The cluster age is determined by fitting the complete cluster sequence to the isochrones models of Bressan et al. (2012). The adopted models were of 1/5\,$Z_{\odot}$ and placed at a $\mu$ of 18.85\,mag corrected for extinction $A_V$\,=\,0.1\,mag. The models spanned from the pre-main sequence (PMS) age of 1\,Myr to an age of 200\,Myr, in logarithmic steps of 0.1\,Myr. The best fit age estimated from the models is 20$^{+12}_{-5}$\,Myr. The non-symmetric error bars take into account the formal fit of the error. The age determination to the cluster sequence does not include the stars identified as H$\alpha$ excess emitters. Including these stars increases the cluster age to 50-80\,Myr. We exclude these stars on the basis that they are presumed cBe stars (concurrent with their positions in the CMD) that canonically fall to the redder side of any given cluster's isochrone, and therefore including them will lead to misleading results.


There is a spread in colour at the low mass end (between $r$ of 22.5--24), where the PMS isochrones of stars older than few Myr arrive on the main sequence. This is most likely the PMS turn-on of the cluster. To the best knowledge of the authors this is the first detection of the PMS turn-on in the Magellanic Bridge, although previous claims of detecting Herbig AeBe stars in and around this particular cluster have been made by Nishiyama et al. (2007), which are discussed in Section\,3.7. These PMS stars lie at the limit of our photometry and show a rather steep detection limit. Deeper photometry from space based instruments may well identify a clear turn-on which is applicable to date the cluster accurately. Comparative near-infrared or X-ray data may well help define the PMS locus of the cluster. It should be noted that despite lying on the PMS tracks, these stars show no H$\alpha$ excess suggesting that they are not presently accreting, in contrast to the findings of a significant population of $\sim$15-25\,Myr {\it accreting} PMS stars detected in metal-poor regions (e.g., see De Marchi et al. 2011 compared to the results from Kalari \& Vink 2015). We estimate that if these stars were strongly accreting with H$\alpha$ line widths $<-$50\AA~in emission, they would fall within the detection limit of our H$\alpha$ photometry suggesting that, at least in NGC\,796, there are no strongly accreting old PMS stars that are similar to those identified in the Magellanic Clouds. This would imply that in the metal-poor diffuse interstellar medium of the Magellanic Bridge, PMS stars lose their discs on timescales younger than $\sim$20\,Myr (e.g. Kalari \& Vink 2015) similar to Galactic PMS stars, and unlike the metal-poor PMS stars found in dense clusters in the Large and Small Magellanic Clouds (e.g. De Marchi et al. 2011) where discs are thought to be retained up\,to ages of 20-25\,Myr. Further deeper observations at infrared wavelengths are essential to verify this.


\subsection{Spectral types}

Spectral classification of IDs\#262 and \#283 (Fig.\,\ref{spectra}) was carried out according to Evans et al. (2004,\,2006), who presented a detailed description of the spectral classification scheme suitable for B-type stars at SMC metallicities. Briefly, the commonly adopted metal lines used to classify B-type stars at SMC metallicities are weaker due to the reduced metallicity. Therefore, our classifications are based primarily on the strength of the Balmer and He\,{\scriptsize I} lines, by visually inspecting the the line strengths, and equivalent widths and comparing them to spectra from Evans et al. (2004, 2006).

In Fig.\,\ref{spectra} we plot the spectra of the NGC\,796 stars, with the main spectral lines identified. Based on the Evans et al. (2004) classification scheme, ID\#283 is classified as B3V, primarily based on the strength of the He\,{\scriptsize I} and Balmer lines, and the absence of strong Si lines (the apparent weak absorption at $\lambda$\,4553\AA~is slightly blue-shifted with respect to the other stellar lines, so appears to be noise rather than Si\,{\scriptsize III} absorption). ID\#262 is classified as a B1V-B3V star. The spectrum is slightly noisier than that of ID\#283, but the strength of the Balmer lines demonstrates sufficiently the classification as an early B1V-B3V star. We also overplot known B2\,V and B3\,V spectral types from high-resolution spectroscopic observations of isolated stars in the SMC cluster NGC\,346 (spectral classifications taken from Bonanos et al. 2010) smoothed to the resolution of our data.

From the spectral types, the effective temperature ($T_{\rm eff}$) of these main-sequence stars can be adequately constrained for further analysis. A spectral type-$T_{\rm eff}$ scale of stars at SMC metallicity is presented in Trundle et al. (2007). Interpolating their results, we estimate that ID\#283 has a $T_{\rm eff}$ $\sim$22500\,K and ID\#262 has a $T_{\rm eff}$ between 27000--22500\,K. The mass ($M_{\ast}$) of these stars can be determined by comparing their estimated $T_{\rm eff}$ to those on the hydrogen burning main-sequence models of Bressan et al. (2012) at 1/5\,$Z_{\odot}$. For ID\#262 we estimate a mass of $\sim$\,11\,$M_{\odot}$, while for ID\#283 we find a mass of $\sim$\,8\,$M_{\odot}$. For ID\#262 (the second brightest star in the cluster without H$\alpha$ excess), the estimated mass translates to a crude main sequence burning lifetime ($\tau_{\rm MS}$) $\sim$ 25\,Myr, which provides an approximate independent upper age limit of the cluster.

Finally, the spectral types can also be used to independently estimate an approximate distance to the cluster, by means of spectroscopic parallax. For ID\,\#262 and ID\,\#283, the $\mu$ estimated from spectroscopic parallax is 18.95 and 18.65\,mag respectively. Both values agree within the errors of the value determined from fitting of the main sequence locus with isochrones. This gives further weight to our determination that the distance to NGC\,796 is 59\,kpc. A summary of the spectral types and estimated spectroscopic parameters are given in Table 3.

\begin{table}
	\centering
	\caption{Summary of spectral types along with the estimated spectroscopic parameters. }
	\label{tab:example_table}
\begin{tabular}{lcccc}        
\hline\hline                
 ID & Sp.\,Type & $T_{\rm eff}$ & $\tau_{\rm MS}$ & $d$  \\
&  & (kK) & (Myr) & (kpc) \\
\hline
\#262 & B1-B3\,V & 27--22.5 & $\sim$25 & 61  \\
\#283 & B3\,V & 22.5 & $\sim$45 & 55  \\
  \hline                        
\end{tabular}\\
\end{table}

\subsection{Comparison with literature}

\begin{figure*}
\centering
\includegraphics[width=1.8\columnwidth]{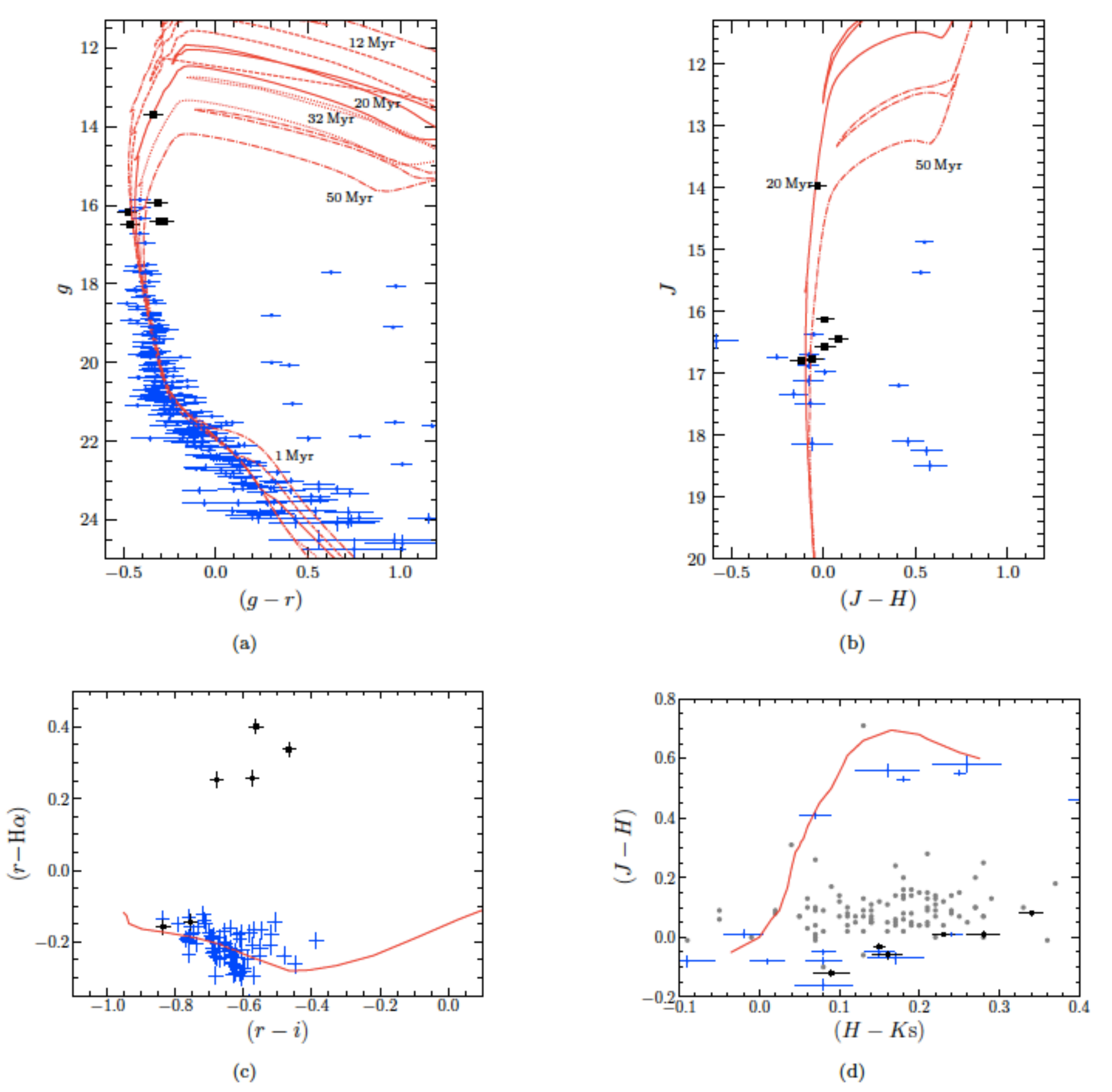}
\caption{(a)($g-r$) vs. $g$ CMD of all stars with optical photometry (blue dots), with the isochrones similar to Fig.\,7. Black asterisks are the Herbig AeBe stars selected by Nishiyama et al. (2007) based on infrared photometry. (b) ($J-H$) vs $J$ CMD of all stars with near-infrared photometry (blue dots) and Herbig AeBe stars from Nishiyama et al. (2007) overplotted with the Bressan et al. (2012) isochrones. (c) ($r-i$) vs. ($r-$H$\alpha$)  TCD with symbols the same as (a). Also shown is the main-sequence locus estimated from the models of Castelli \& Kurucz (2003). (d) ($H-K$s) vs. ($J-H$) TCD with symbols same as (b). Overplotted is the near-infrared locus from Bessell \& Brett (1988) transformed for the IRSF filters. Grey circles are the positions of cBe stars in the SMC from Martayan et al. (2010). }
\label{cmdall2}
\end{figure*}

Three optical and one near-infrared study of NGC\,796 are found in the literature. Ahumada et al. (2002,2009) used integrated optical spectra to determine the cluster properties. They fit the observed spectra to templates to determine the reddening and age by comparing the equivalent widths of Balmer and metal absorption lines. For NGC\,796 they determine a reddening $E(B-V)$ between 0.03--0.06, and an age of 20$\pm$10Myr. They also find moderate emission in H$\alpha$, which they attribute to cBe stars. Our results agree well with those from Ahumada et al. (2009).

Piatti et al. (2007) utilized two-colour Washington photometry to determine the cluster properties (given as L115 in that paper). They also determined the cluster reddening by interpolating H\,{\scriptsize I} and 100\,$\mu$m dust emission maps, finding a $E(B-V)$ of 0.03. By fitting theoretical isochrones to the cluster main sequence, they estimate a cluster age of 110$^{+50}_{-20}$\,Myr at an adopted $\mu$ of 18.77\,mag. The age determined by Piatti et al. (2007) is much older than the age determined in this paper and by Ahumada (2009), although the reddening agrees within errors. On inspection of their data, we find that the brightest stars are most likely saturated and therefore appear redder, as their shortest reported exposure time is 1600s on a 1.2m telescope, at the limit where stars brighter than $V$\,$<$\,17 should become saturated. Another recent study by Bica et al. (2015) also arrived at the same conclusion. Also, inspecting their CMD it is apparent that they fit isochrones to the cBe stars which are redder than those on the main sequence and whose inclusion might lead to an incorrect age estimate. 

Bica et al. (2015) utilized $BV$ photometry to determine the cluster properties by fitting the cluster locus with theoretical isochrones. They quoted multiple ages and distances for  NGC\,796 at different metallicity assumptions ($Z$\,=\,0.01 and 0.006), finding $E(B-V)$ to be between 0.03--0.04 for both cases. However, in the $Z$\,=\,0.01 case they find a slightly older age of 42$^{+24}_{-15}$\,Myr for the cluster but at a much smaller distance of 40.6\,kpc. Similarly, in the $Z$\,=\,0.006 case (which is closer to the metallicity assumed in this work), they report an age of 16$^{+6}_{-5}$\,Myr at a distance of 39.2\,kpc. Such distances would indicate that the Bridge is much closer to us than both the LMC and SMC, contrary to most previous literature determinations of the distance to the Magellanic Bridge (D'Onghia \& Fox 2016). Further more, the spectroscopic parallax provides a further confirmation that the distance to NGC\,796 is similar to the SMC, rather than in front of the LMC. 


Nishiyama et al. (2007) utilized near-infrared {\it JHK}s photometry of the Infrared Survey Facility Survey (IRSF; Kato et al. 2008) to identify Herbig AeBe candidates in the cluster. Their findings are based on the positions of the stars in the ($J-H$) vs. ($H-K$s) TCD. We cross-matched our photometry with the IRSF data, and plot the resulting TCD and CMD in both optical and near-infrared filters in Fig.\,\ref{cmdall2}. The cross-matched stars are reported with their IRSF identifications in the final column of Table 2. In Fig.\,\ref{cmdall2}a,b we show the ($g-r$) vs. $g$ and ($J-H$) vs. $J$ CMD of all stars with optical and near-infrared photometry, respectively. The stars identified by Nishiyama et al. (2007) as Herbig AeBe stars are marked. Comparing their positions, they appear to be likely Be candidate stars. We further inspect this by plotting them in the ($r-i$) vs. ($r-$H$\alpha$) and ($J-H$) vs. ($H-K$s) TCD in Fig.\,\ref{cmdall2}c,d. Also shown are the positions of known SMC cBe stars from the survey by Martayan et al. (2010). On inspecting the TCDs, it is apparent the Herbig AeBe stars found by Nishiyama et al. (2007) are most likely cBe stars, i.e. evolved stars with H$\alpha$ emission and near-infrared excesses. They have colours corresponding to early-type stars and their H$\alpha$ excesses correspond to H$\alpha$ equivalent widths between $-$5 to $-$30\AA ~(compared to the TCD in Barentsen et al. 2011), in keeping with the equivalent widths for SMC cBe stars reported by Martayan et al. (2010). Their near-infrared excesses are also in keeping with those reported in Martayan et al. (2010), whereas Herbig AeBe stars have much larger ($H-K$s) excesses, even at lower metallicities.

\section{Discussion}

\subsection{Initial mass function}

\begin{figure}
\centering
\includegraphics[width=0.9\columnwidth]{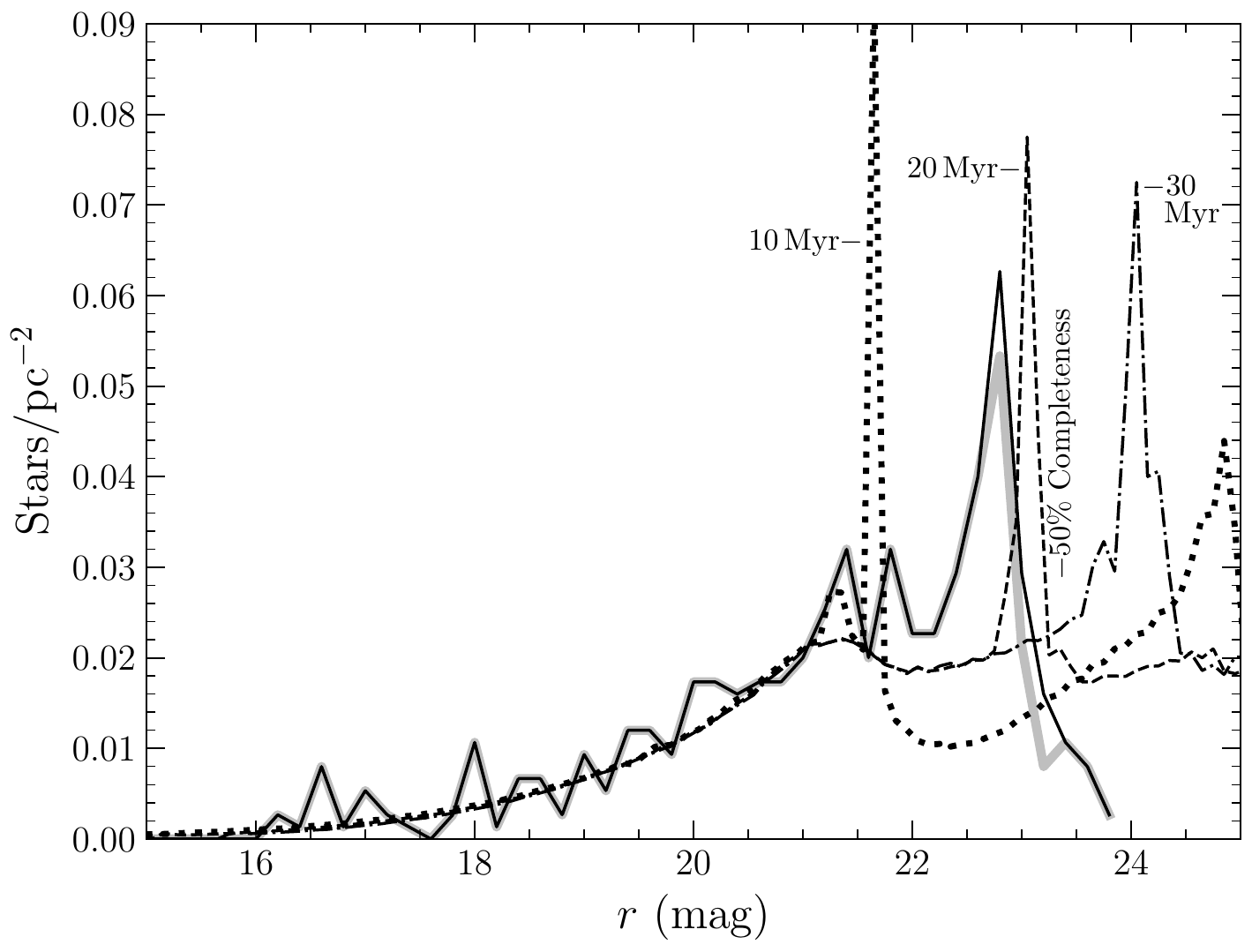}
\caption{The $r$-band luminosity function of NGC\,796 normalized per unit area and corrected for incompleteness is given by the solid black line, while the grey line is the luminosity function uncorrected for completeness. Overlaid as dotted (leftmost), dashed and dashed dotted (rightmost) are the Bressan et al. (2012) luminosity functions of 1/5\,$Z_{\odot}$ calculated assuming a Salpeter IMF of stars at 10, 20 and 30 \,Myr, respectively. The 50\% completeness limit is marked. \label{lumfun}}
\vspace{-0.7em}
\end{figure}

In Fig.\,\ref{lumfun}, we show the $r$-band luminosity function of the NGC\,796 cluster members corrected for incompleteness based on the discussion in Section\,2.1. Also shown is the luminosity function uncorrected for incompletness, with the 50\% completeness limit of the photometry marked. We only plotted stars that were also detected in $i$-band photometry, and classified as cluster members based on their $r_{\rm t}$, and position in the CMD. The luminosity function is corrected for extinction, but not distance. Overplotted are the luminosity functions (derived assuming a Salpeter stellar IMF) from 10, 20 and 30\,Myr isochrones from Bressan et al. (2012). There is a clear PMS turn-on identified in the cluster luminosity function, which demonstrates the result obtained in Section\,3.5.

The luminosity function can be translated into a mass function assuming a mass-luminosity relation. The derived mass function is representative of the cluster mass function at the present day, i.e. the present-day mass function (PDMF). The IMF can then be estimated from the PDMF assuming a star-formation history (Elmegreen \& Scalo 2006), and also by accounting for stars that have ended their lives. We assume a single burst of star formation at 20\,Myr. Since we do not know the mass of the most massive star born in the cluster, we cannot accurately estimate the mass function of stars that have already ended their lives. Instead we limit our PDMF determination to 10\,$M_{\odot}$, the mass in the Bressan et al. (2012) models at which a star born 20\,Myr ago is still on the hydrogen burning main sequence. Then, the PDMF for $M_{\ast}<$10\,M$_{\odot}$ accurately represents the IMF of the cluster within that mass range (see Elmegreen \& Scalo 2006).

\begin{figure}
\centering
\includegraphics[width=0.9\columnwidth]{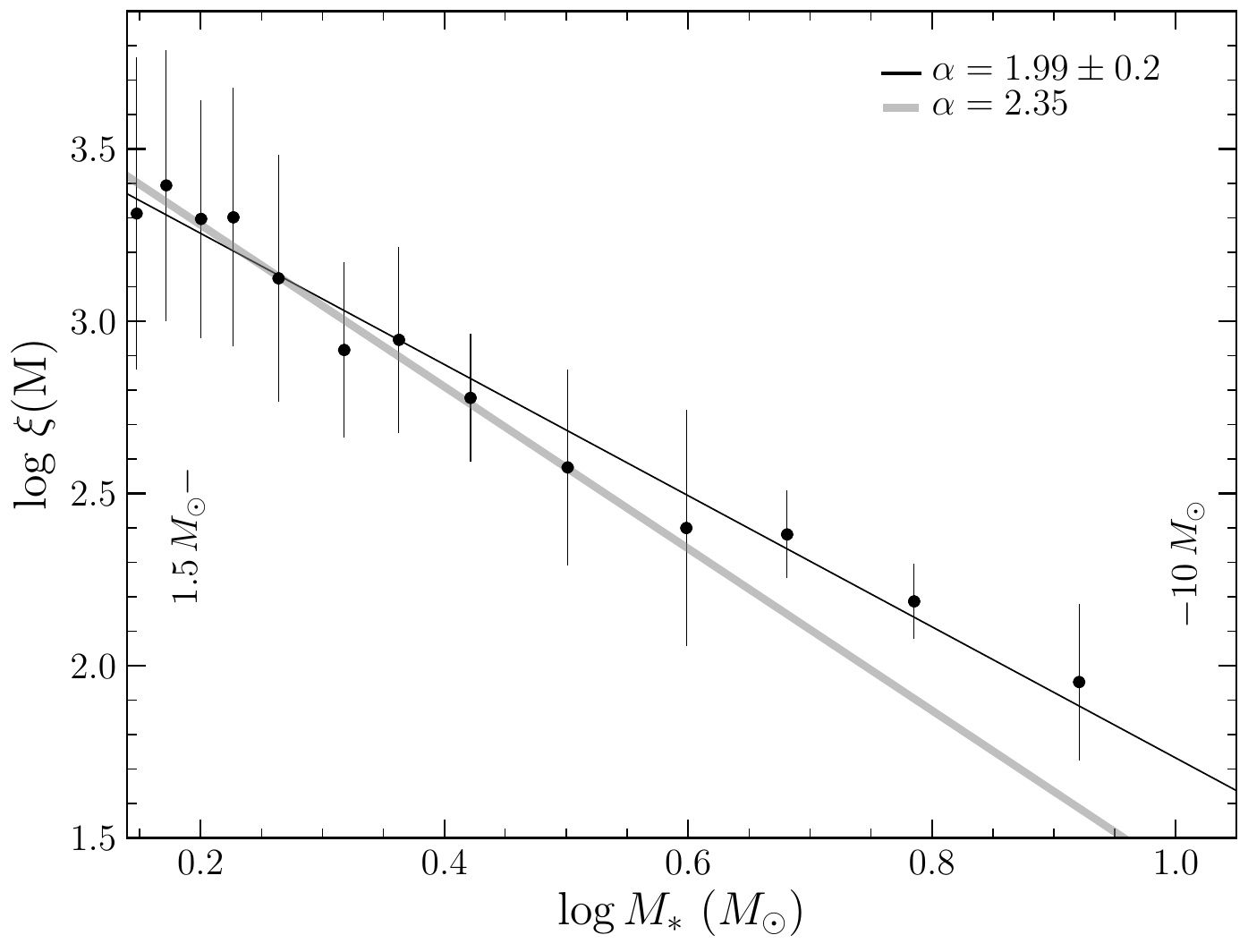}
\caption{Stellar IMF of NGC\,796 in logarithmic mass, with the best-fit slope of $\alpha$\,=\,1.99 for the 1.5-10\,$M_{\odot}$ range. The slope for a Salpeter IMF ($\alpha$\,=\,2.35) is also shown. \label{imf}}
\end{figure}

To derive the IMF we use the mass-luminosity relation of the 1/5\,$Z_{\odot}$ 20\,Myr old isochrone from Bressan et al. (2012). We estimated errors on the IMF by shifting the mass-luminosity relation by 0.3\,mag to account for the combined error of distance and isochronal age estimates, propagating them into the error bars on our PDMF. We limit our investigation to the mass range wherein we are complete, i.e. $>$1.5\,$M_{\odot}$ which is the 50\% completeness limit. This limit covers stars within the mass range of 10\,$M_{\odot}$--1.5\,$M_{\odot}$. Following Ma{\'i}z Apell{\'a}niz \& Ubeda (2005), we adopt uniform quantiles of unit one to derive the mass function. This method is preferred over conventional equispaced binning as assuming a constant bin size may lead to misleading correlations between the number of stars per bin along the mass range. The authors of that paper demonstrate that with variable bin size, but keeping the number of sources per bin constant limits the errors in the low number statistics regime. This is particularly important for deriving mass functions at the high mass end, where the number of stars is smaller compared to the low mass end. The mass function estimated this way is shown in Fig.\,\ref{imf}. Using a linear regression fit to the IMF, we find a slope $\alpha$ of 1.99$\pm0.2$. Here, $\alpha$ is the IMF slope when written in linear mass units as, 
\begin{equation}
\chi(m) = dN/dm \propto m^{-\alpha}
\end{equation}

The resulting value is 1$\sigma$ smaller than the canonical Salpeter IMF, that has a value of $\alpha$\,=\,2.35 (Bastian et al. 2010) which has been reported for the majority of open clusters in the SMC (see the review by Bastian et al. 2010). Significantly, it is effectively at the critical value of the IMF (at $\alpha$\,=\,2) where there is more mass at higher rather than lower mass, i.e. a slightly top-heavy IMF. It has been suggested that the IMF of metal-poor regions is likely to be more top heavy. Consider that when molecular clouds collapse to form stars, their gravitational energy is converted to thermal energy. In nearby metal-rich molecular clouds thermal energy is dissipated through metal line emission and dust cooling, advancing fragmentation and leading to the low characteristic stellar mass (0.3\,$M_{\odot}$), of the present-day IMF (Kroupa 2002). In contrast, the lack of cooling due to the absence of metals, and low dust content is thought to inhibit fragmentation resulting in a higher characteristic stellar mass at lower $Z$ (potentially up to $\sim$ 100\,$M_{\odot}$ for the metal devoid Population III stars). This is generally argued to result in a top-heavy IMF in the early Universe (e.g. Bromm \& Larson 2004). Thus, our result is consistent with suggestions that the IMF is more top heavy at lower metallicities. The top-heavy IMF could also be an environmental effect owing to the diffuse interstellar medium of the Bridge. In either scenario, by extension of the top heavy IMF to star formation at the low-metallicities ($<$1/10\,$Z_{\odot}$) expected in the early Universe, this would lead to a greater number of high-mass stars in distant star-forming galaxies than we see in the Milky Way today, impacting on properties such as ionizing fluxes and stellar feedback.

However, we caution that no conclusive evidence for a top heavy IMF in open clusters in the SMC and LMC as a function of $Z$ has yet been put forward (Bastian et al. 2010), and theoretically it is unclear at what critical $Z$ the IMF begins to show observable changes or reaches the critical value between top or bottom-heavy, or whether the observed changes are an environmental effect. While it maybe possible that the $Z$ of NGC\,796 is much lower than the SMC at around 1/10\,$Z_{\odot}$, we refrain from drawing strong conclusions based on our result. Deeper optical/infrared observations of the low-mass stellar population; and spectroscopic estimates of the upper IMF (e.g. Schneider et al. 2018) are essential to validate this result. Finally, adopting the Kroupa (2001) IMF, where the IMF slope at masses upwards of 0.5\,$M_{\odot}$ is constant, we can estimate the total cluster mass between 0.5\,$M_{\odot}$--10\,$M_{\odot}$. Adopting $\alpha$\,=\,1.99, the total cluster mass ($M_{\rm cl}$) between these limits is 990$\pm$200\,$M_{\odot}$.

\subsection{Be candidate stars}

The Be phenomenon, where a main-sequence B-type star displays H$\alpha$ emission is thought to arise due to matter ejections from the central star forming a rotationally supported disc. These stars are presumed to rotate at speeds very close to their critical rotational velocities and are near breakup. Maeder et al. (1999) estimate the fraction of cBe stars relative to all known B spectral type stars in 21 clusters in the SMC (at 1/5\,$Z_{\odot}$), LMC (1/2\,$Z_{\odot}$) and the Galaxy. They found that the fraction of cBe stars increases with decreasing metallicity (although they only studied one cluster in the SMC). The results of Maeder et al. (1999) advocate the possibility of faster rotation at lower metallicities, which is a key component of the chemically-homogeneous evolution theory at lower metallicities. Further studies by Martayan et al. (2010) including multiple SMC and Galactic clusters found that although the average frequency of cBe to B stars increases as a function of $Z$, there are further degeneracies due to the cluster age, and possibly other factors.

In Fig.\,\ref{ccd} the ($r-i$) colour of members is plotted against the ($r-$H$\alpha$) colour, which is a measure of the strength of H$\alpha$ line emission, from which the H$\alpha$ emitters are easily identified. On restricting our magnitude range in $r$ to only include main-sequence B0-B3 stars, we can quantify the fraction of early type B stars with H$\alpha$ excess as a fraction of the total B stars in NGC\,796. Given the cluster's age of $\sim$ 20\,Myr, the amount of H$\alpha$ excess and the correlated near-infrared ($H-K$s) excess, we are confident that these stars are cBe stars. The ratio of cBe stars to (cBe+B) stars ($R_{\rm cBe}$) in the B0-B3 spectral range was determined to be 37$\pm$6\%, where the error is the propagated uncertainty in the photometric classification of B spectral types. Considering all B spectral type stars (assuming apparent $r<19$ considering extinction), $R_{\rm cBe}$ drops to 31$\pm$7\%.  Given that the H$\alpha$ selection criterion is only an upper limit, the actual $R_{\rm cBe}$ may be higher. The values determined here reflect only a single cluster, but form an important data point because they do not suffer from the various issues such as variable extinction, cluster membership and crowding faced by similar studies in the Magellanic cloud clusters (e.g. Evans et al. 2006).

When comparing our results with those of Maeder et al. (1999), we adopt their absolute magnitude cuts, assuming $M_V$$\sim$$M_r$. For their only studied SMC cluster, NGC\,330 (at 20\,Myr) they report a $R_{\rm cBe}$ of 39\% in the $-5<M_{V}<-1.4$ range, 44\%  in the $-5<M_{V}<-2$ range, and 46\% in the $-4<M_{V}<-2$ range. Within the same absolute magnitude ranges, we find a $R_{\rm cBe}$ of 47\%, 50\%, and 45\%, respectively. Martayan et al. (2010) found that at an age of 20\,Myr, the average $R_{\rm cBe}$ at the SMC $Z$ at B0-B3 spectral types is $\sim$10\%, considerably lower than our derived estimate. If the metallicity of NGC\,796 were lower, it would extend the $Z$ range of such relations; however future spectroscopy is required both to determine stellar metallicies and verify the candidate emission-line stars.

\subsection{Massive open cluster}

\begin{figure}
\centering
\includegraphics[width=0.9\columnwidth]{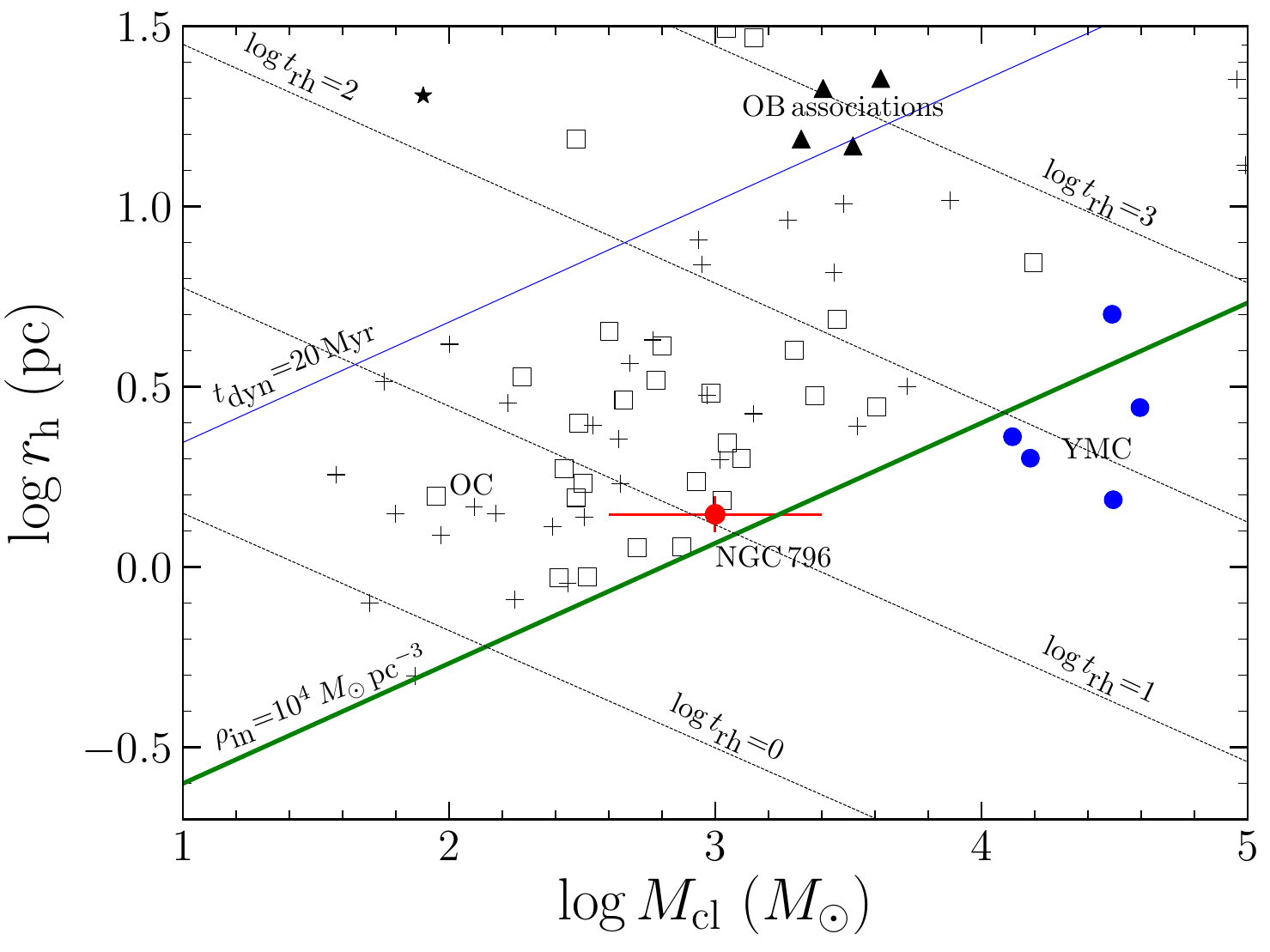}
\caption{Half mass radius vs. cluster mass diagram. NGC\,796 (red circle) is plotted in comparison with 20 Myr open clusters (crosses), 20 Myr young massive clusters (YMCs; blue circles), and OB associations (triangles) from the literature (see text for details). The black asterisks denotes clusters in the Magellanic Bridge from Mackey et al. (2017) younger than 100\,Myr; and the black squares SMC open clusters younger than 100\,Myr from Hunter et al. (2003) and Maia et al. (2014). The green line denotes a cluster with initial density greater than 10$^4\,M_{\odot}\,$pc$^{-3}$, and the blue line shows the dynamical timescale of 20\,Myr. Lines of equal relaxation time are also shown in solid black.  \label{mcl}}
\end{figure}

To classify NGC\,796 within the context of known clusters, we place it in the half-mass radius ($r_{\rm h}$) vs. $M_{\rm cl}$ diagram of Portegies Zwart et al. (2010) in Fig.\,\ref{mcl}. Here, $r_{\rm h}$ can be considered to be the $r_{\rm c}$ of the cluster. Portegies Zwart et al. (2010) and Fujji \& Portegies Zwart (2016) compared the distribution of Galactic open clusters (OC), OB associations, and young massive clusters (YMC) in the $r_{\rm h}$ vs. $M_{\rm cl}$ diagram. They suggested that OC and OB associations follow different evolutionary paths to YMCs. We can classify NGC\,796 by comparing its the position in Fig.\,\ref{mcl} to the literature clusters, with it best described as a massive open cluster. It is more compact than previously known clusters in the Magellanic Bridge (Mackey et al. 2017), and as compact and massive as those known in the SMC (Hunter et al. 2003; Maia et al. 2014). It closely follows the evolutionary track for massive clusters given by Fujji \& Portegies Zwart (2016) of $r_{\rm h}$ (pc)\,=\,0.34\,$\times$\,Age$^{3/2}$ (Myr). On comparison, it is also clear that NGC\,796 is not expected to become unbound soon, lying below the relation where the cluster's age is equal to its dynamical age (above which clusters dissolve into the field). It is also near the relation where clusters with initial densities of $\rho$\,=\,10$^4 M_{\odot}\,$pc$^{-3}$ formed, therefore it is a 20\,Myr dense massive open cluster located in the Magellanic Bridge.

It is extremely interesting to consider how NGC\,796 formed as it is challenging to explain within the current competing models of in-situ or conveyor belt star-formation for massive open clusters given the low surface density and volume of molecular gas (Mizuno et al. 2006). Almost all other clusters in the Magellanic Bridge are loose associations at typical radii of a few tens of pc (e.g. Mackey et al. 2017), or low-mass clusters with masses less than 100\,$M_{\odot}$ (e.g. Bica \& Schmitt 1995), making NGC\,796 unique in this respect as a massive open cluster. The surrounding molecular clouds have masses between 1-7$\times$10$^3$$M_{\odot}$, with the nearest molecular cloud 60\,pc to the south having a mass of 7$\times$10$^3$\,$M_{\odot}$. If we consider NGC\,796 formed from a cloud of similar mass, it must have star-formation efficiencies around $\sim$10\%, which is much higher than that of most known star-forming regions (Kennicutt \& Evans 2012), suggesting the parent cloud must have been much larger. In this scenario the molecular cloud is formed from native H\,{\scriptsize I} gas in the Bridge (which is thought to have been created by an encounter between the LMC and SMC according to Besla et al. 2012; Gardiner et al. 1994). This initial cloud made of native Bridge molecular material in time grew and collapsed to form NGC\,796, either monolithically from a massive molecular cloud, or hierarchically through the collapse of smaller Bridge clouds. If this scenario is correct, NGC\,796 is an ideal laboratory for future studies to test star-formation models at their current extremities, and in an low-$Z$ environment similar to that found in the early Universe. An alternative explanation could be that NGC\,796 was ejected after forming in the wing of the SMC. Although, in this scenario the cluster would be moving away from the SMC at a velocity of $\sim$200\,km\,s$^{-1}$ (as it is located 4000\,pc from the wing at an age of 20\,Myr). Future measurements of the velocity are necessary to rule out this possibility, but we consider it unlikely given the lack of a an ejection mechanism (e.g. similar to the twin supermassive black holes in the ejected globular cluster HVGC-1; Caldwell et al. 2014).


 
\section{Conclusions}

We present deep AO {\it gri}H$\alpha$ photometry of the cluster NGC\,796 in the Magellanic Bridge, which we assume has a $Z$ of 1/5\,$Z_{\odot}$. From the position of the zero age main sequence locus in the two-colour and colour-magnitude diagrams we constrain the cluster's visual extinction $A_V$ to be 0.1\,mag, and find it is located at a distance modulus of 18.85$\pm$0.2\,mag, corresponding to a distance of 59$\pm$0.8\,kpc, placing it in between the SMC ($\sim$ 61\,kpc) and the LMC (50\,kpc). In the cluster's CMD and luminosity function, we identify a pre-main sequence turn-on at r$\sim$23 (or around 1.5\,$M_{\odot}$). From fitting isochrones to the cluster locus we determine an age of 20$^{+12}_{-5}$\,Myr. The photometric estimates of distance and age compare well with the spectroscopic parallax distance estimates and main sequence burning lifetime of two bright cluster members, which were classified as early B-type stars.

Based on a theoretical mass luminosity relation, we estimated the IMF of the cluster. The slope of the IMF when written in linear mass units is $\alpha$\,=\,1.99$\pm0.2$. We argue that the derived slope may hint at a metallicity (or environmental) effect on the IMF (cf. the Salpeter IMF with $\alpha$\,=\,2.35 determined in the majority of clusters in the SMC as summarized by Bastian et al. 2010).

From the ($r-$H$\alpha$) colours of our stars, we identify cBe stars and compute the relative fraction of cBe stars to total (cBe+B) stars within defined spectral type and absolute magnitude ranges. Within the uncertainties, this ratio is comparable to that for NGC\,330 in the SMC, the largest Be-fraction known to date.

Finally, based on the derived cluster properties we classify NGC\,796 as a massive open cluster, and speculate about its formation in the diffuse interstellar medium of the Magellanic Bridge. We suggest that current star and cluster formation theories, particularly those at low metallicities (or high redshifts) may utilize the derived properties of NGC\,796 to constrain their models.

\acknowledgments

We thank the anonymous referee for a detailed, constructive and very helpful report. This paper made use of data obtained under the Chilean National Telescope Allocation Committee time under the programme IDs.\,38317833 \& 12990104. Based on observations obtained at the Southern Astrophysical Research (SOAR) telescope, which is a joint project of the Ministerio da Ciencia, Tecnologia, e Inovacao (MCTI) da Republica Federativa do Brasil, the U.S. National Optical Astronomy Observatory (NOAO), the University of North Carolina at Chapel Hill (UNC), and Michigan State University (MSU). V.M.K. thanks Dr. Jashty Ashalata for comments on an earlier draft, and Drs. Leonardo Kebler \& Andres Piatti for useful discussions. V.M.K. acknowledges Leonardo Kebler, Sergio Pizarro, and Andrei Tokovin for their assistance during the SOAR observing run. V.M.K. also thanks Nidia Morrel, and the Las Campanas staff for their assistance in using the Bollers \& Chivens Spectrograph, and the hospitality and support provided during the LCO runs. V.M.K. acknowledges support from the FONDECYT-CHILE Fellowship grant N$^{\rm o}$.\,3116017. M.R. acknowledges support from the FONDECYT-CONICYT grant N$^{\rm o}$.\,1140839 \& BASAL\,PFB-06.

%

\vspace{5mm}

\end{document}